\documentclass[twoside,english]{elsarticle}
\usepackage[T1]{fontenc}
\usepackage[latin9]{inputenc}
\usepackage{array}
\usepackage{float}
\usepackage{multirow}
\usepackage{amsbsy}
\usepackage{amstext}
\usepackage{amsthm}
\usepackage{graphicx}
\usepackage{esint}
\usepackage{url}
\usepackage{epstopdf}

\makeatletter

\providecommand{\tabularnewline}{\\}
\newcommand{\lyxdot}{.}

\theoremstyle{plain}
\newtheorem{thm}{\protect\theoremname}[section]
\theoremstyle{definition}
\newtheorem{defn}[thm]{\protect\definitionname}
\theoremstyle{plain}
\newtheorem{cor}[thm]{\protect\corollaryname}
\ifx\proof\undefined
\newenvironment{proof}[1][\protect\proofname]{\par
\normalfont\topsep6\p@\@plus6\p@\relax
\trivlist
\itemindent\parindent
\item[\hskip\labelsep\scshape #1]\ignorespaces
}{%
\endtrivlist\@endpefalse
}
\providecommand{\proofname}{Proof}
\fi

\journal{Electric Power Systems Research}

\makeatother

\usepackage{babel}
\providecommand{\corollaryname}{Corollary}
\providecommand{\definitionname}{Definition}
\providecommand{\theoremname}{Theorem}

\begin{document}

\begin{frontmatter}{}

\title{Rational consumer decisions in a peak time rebate program}

\author[focal]{José~Vuelvas\fnref{fn1}}

\author[focal]{Fredy~Ruiz\fnref{fn2}}

\fntext[fn1]{vuelvasj@javeriana.edu.co.}

\fntext[fn2]{ruizf@javeriana.edu.co.}

\address[focal]{Pontificia Universidad Javeriana, Bogotá, Colombia}
\begin{abstract}
A rational behavior of a consumer is analyzed when the user participates
in a Peak Time Rebate (PTR) mechanism, which is a demand response
(DR) incentive program based on a baseline. A multi-stage stochastic
programming is proposed from the demand side in order to understand
the rational decisions. The consumer preferences are modeled as a
risk-averse function under additive uncertainty. The user chooses
the optimal consumption profile to maximize his economic benefits
for each period. The stochastic optimization problem is solved backward
in time. A particular situation is developed when the System Operator (SO)
uses consumption of the previous interval as the household-specific
baseline for the DR program. It is found that a rational consumer
alters the baseline in order to increase the well-being when there
is an economic incentive. As results, whether the incentive is lower
than the retail price, the user shifts his load requirement to the
baseline setting period. On the other hand, if the incentive is greater than
the regular energy price, the optimal decision is that the user spends
the maximum possible energy in the baseline setting period and reduces
the consumption at the PTR time. This consumer behavior produces more
energy consumption in total considering all periods. In addition,
the user with high uncertainty level in his energy pattern should
spend less energy than a predictable consumer when the incentive is
lower than the retail price. \end{abstract}

\end{frontmatter}{}

\section{Introduction}

In the smart grid concept, DR is a mechanism implemented by SO to
equilibrate the load with power generation by modifying consumption.
The main purpose of this kind of program is to curtail load at the
peak demand times for maintaining the security of the transmission
assets, avoiding to exceed the limit capacity of generators and preventing
power outages. Therefore, DR is one of the most crucial parts of the
future smart grid \citep{Zhu2013} due to the main objectives of the
DR is peak clipping, valley filling and load shifting on the power
profile. An important question in DR program design is how to improve
the demand profile, namely, to control the noncritical loads at the
residential, commercial and industrial levels for matching supply
and demand. For instance, DR program might motivate changes in electricity
usage by changing the price of electricity or giving an incentive
payment. 

There are several DR programs implemented as part of strategies to
reduce peak power (because the demand trend is growing). In \citep{Vardakas2015,Deng2015}
are shown a complete summary regarding mathematical models, pricing
methods, optimization formulation and future extensions. The common
approach is time-varying pricing (TVP), which charge more money for
energy use during peak periods. In TVP program, the consumer does
not have a significant incentive to curtail the consumption, just
the energy is more expensive at certain hours. Others mechanisms have
been implemented where the user behavior is modified through economic
incentives, therefore, many utilities have employed a change in the
residential electricity rate structure \citep{Newsham2010}. For instance,
Time-of-use (TOU) \citep{Datchanamoorthy2011} program, where the
day is divided into adjoining blocks of hours. The price of energy
varies between blocks, but not within blocks; Critical peak pricing
(CPP) \citep{Herter2007}, is related to TOU, unlike that it is only
applied to a small number of event days; in Real-time pricing (RTP)
\citep{Bloustein2005}, the price varies hourly according to the real-time
market cost of delivering electricity; Direct load control \citep{Ericson2009},
remote control of flexible loads; Emergency demand reduction \citep{Tyagi2010},
users receive incentive by diminishing energy consumption during emergency
events; PTR \citep{Vuelvas2015}, where customers receive electricity
bill rebates by not consuming (relative to a previously established,
household-specific baseline) during peak periods, which it is the
mechanism studied in this paper; and many other mechanisms. 

The baseline is an important concept of the PTR program. A counterfactual
model is developed to estimate the baseline. In \citep{Chao2011}
shows the critical facts on the selection of customer baseline, they
design a suitable baseline focusing on administrative and contractual
approaches in order to get an efficient DR. Furthermore, in \citep{Faria2013,Wijaya2014,Antunes2013}
the performance of DR baselines are studied and new methods are regarded
as establishing the reasonable compensation of the consumer. 

Moreover, in literature, there are some of more theoretical DR programs
such as in \citep{Zhong2012} by using a smart grid technology, the
authors shows a DR program where a load device could offer retail
users coupon incentives to induce DR for a future period in anticipation
of intermittent generation. In \citep{Huang2015} a cooperative dynamic
DR under different market architecture is proposed to evaluate the
welfare impacts and the efficiency-risk. In addition, \citep{Nguyen2013}
devise schemes for scheduling DR in a deregulated environment. The
authors create a new market concept trough a pool-based market-clearing
strategy. In \citep{6161320} a real-time DR algorithm is developed.
Furthermore, it is possible to find DR program based on the game theory
such as \citep{Nekouei2015} to ensure that users tell the truth in
relation to their reduced power consumption employing a Vicrey-Clarke-Grove
mechanism. 

In this paper, a rational consumer behavior is studied when he is
enrolled in a particular DR incentive program based on baseline or
counterfactual model called Peak Time Rebate (PTR). In the economic
sense, rational behavior means that the users maximize their profits
given the mechanism of demand energy reduction. This rebate is calculated
using a baseline for each user which is estimated from past energy
consumption. In real life, the PTR program has shown to be an inefficient
DR mechanism to improve the demand profile because it allows that
some users deliberately increase consumption during baseline-setting
times \citep{Wolak2006,SeverinBorenstein2014,LLP}. Such consumer
behavior of altering the baseline is formulated as a stochastic optimization
problem to understand how the users take their decisions of consumption
when they are participating in PTR program. While the user intuitively
makes decisions according to the operation of the mechanism, in this
work, a mathematical model of consumer choice is proposed in order
to find solutions to the aforementioned inefficiency of the PTR mechanism.

In this work, the optimal strategy of a user that participates in
a PTR program is studied in order to earn the highest economic profit
under uncertain decisions.The contribution is described as follows:
\begin{itemize}
\item The optimal decision problem is posed in general form taking into
account several previous periods of setting-time in a PTR program.
The purposed solution is solved backward in time to find the optimal
choice for consumers where consumer uncertainty is modeled as a random
variable. In addition, the choice of the SO is modeled as a binary
random variable, namely, for indicating whether the user is called
for participating in PTR mechanism. 
\item A closed form solution of a PTR program is derived for two periods.
The previous consumption is assumed as the baseline and the user is
always called to participate in the PTR program. The results show
that the consumer alters the baseline when the incentive exists in
the DR program. Some numerical examples are presented.
\end{itemize}
The article is organized as follows. Section II describes the preliminary
setting. In Section III, the general problem formulation of the PTR
program is developed. Section IV, the mathematical solution for two
periods given the optimization problem is explained. Section V, the
simulation results are shown. Conclusions are presented in Section
VI.

\section{Setting}

This section presents the notation and assumptions for developing
the model. An individual consumer or aggregated demand (a group of
users with the same or similar preferences) is considered for this
DR model. the decision maker's preferences are specified by giving
utility function $G(q_{t};\theta_{t})$, where $q_{t}$ is the consumption
at time $t$ and $\theta_{t}$ is a particular realization of random
variable $\Theta$. The randomness $\Theta$ are external factors
that influence the energy requirements of the consumer. The randomness
in the utility function is modeled as an additive load requirement,
that is, $G(q_{t};\theta_{t})=G(q_{t}-\theta_{t})$. $\Theta$ is
assumed to have a probability density function $f_{\Theta}(\theta_{t})$
with limited support $\left[\underline{\theta},\overline{\theta}\right]$
and mean zero. The motivation to choose such additive randomness is
that an external event, such an as a cold wave, will drive the user
to increase his energy consumption until he obtains the same comfort
than without the event. Then, given a price, the effect of the random
event is to shift the equilibrium point to the left in this situation.

The consumer is assumed with risk-averse behavior. Individuals will
usually choose with lower risk, therefore, $G(\cdotp)$ is concave
\citep{VegaRedondo2003}. This behavior reflects the assumption that
marginal utility diminishes as wealth increases. Also, $G(\cdotp)$
is considered smooth, positive and nondecreasing.

A competitive electricity market (consumers are price-takers) is assumed.
Thus, the energy price $p$ is given and constant since the utility
company set an invariable price to the users during the certain period.
Then, the following definitions are stated.
\begin{defn}
The energy total cost is $\pi(q_{t})=pq_{t}$. 
\end{defn}

\begin{defn}
The payoff function is defined as $U_{t}\left(q_{t},\theta_{t}\right)=G(q_{t}-\theta_{t})-\pi(q_{t})$,
which indicates the user benefit of consuming $q$ energy during the
interval $t$. 
\end{defn}

\begin{defn}
Given $G(\cdotp)$, $\theta_{t}$ and $p$, the rational behavior
of the consumer that maximizes the payoff function $U_{t}\left(q_{t},\theta_{t}\right)$
is 
\begin{equation}
q_{t}^{*}(\theta_{t})=\overline{q}+\theta_{t}\label{eq:q*}
\end{equation}
this result is found by solving the optimization problem

\[
q_{t}^{*}=\max_{q_{t}\in\left[0,q_{max}\right]}\:U_{t}\left(q_{t},\theta_{t}\right)=G\left(q_{t}-\theta_{t}\right)-\pi\left(q_{t}\right)
\]

where $q_{max}$ is the maximum allowable consumption value, $\overline{q}$
is the optimal solution to the previous condition when $\theta_{t}=0$. 
\end{defn}

\subsection{Utility function and rebate description}

Under assumption that $G(\cdotp)$ is a smooth and concave function,
the utility function can be approximated by a second order polynomial
around $\overline{q}$. Therefore, a quadratic function is considered,
where the user utility is zero whether his consumption is zero and
saturates after achieving the maximum of the quadratic form, i.,e.,

\[
G\left(q_{t}\right)=\left\{ \begin{array}{cc}
-\frac{\gamma}{2}\left(q_{t}-\overline{q}\right)^{2}+p\left(q_{t}-\overline{q}\right)+k & \quad0\leq q_{t}\leq\overline{q}+\frac{p}{\gamma}\\
-\frac{p^{2}}{2\gamma}+\frac{p^{2}}{\gamma}+k & \quad q_{t}>\overline{q}+\frac{p}{\gamma}
\end{array}\right.
\]

The saturated part is motivated due to the fact that the agent has
a limited well-being with respect to his energy consumption.
\begin{defn}
Under additive uncertainty and using the previous consideration (\ref{eq:q*}),
The utility function can be rewritten as follow:

\begin{equation}
G\left(q_{t}-\theta_{t}\right)=\left\{ \begin{array}{cc}
-\frac{\gamma}{2}\left(q_{t}-q_{t}^{*}\right)^{2}+p\left(q_{t}-q_{t}^{*}\right)+k & \quad0\leq q_{t}\leq q_{t}^{*}+\frac{p}{\gamma}\\
-\frac{p^{2}}{2\gamma}+\frac{p^{2}}{\gamma}+k & \quad q_{t}>q_{t}^{*}+\frac{p}{\gamma}
\end{array}\right.\label{eq:g}
\end{equation}
\end{defn}

where $\gamma$ and $k$ are constant. In particular, $\gamma$ depicts
consumer private preferences and $k$ is settled when $G\left(q_{t}-\theta_{t}\right)=0$
if $q_{t}-\theta_{t}=0$. A similar approach to model a utility function is found in \cite{6266724}. A further discussion about $\gamma$ can be reviewed in \cite{Fahrioglu2001}

Note that $\gamma$ is in dollar or any other currency divided by
energy units squared, therefore, this parameter could be interpreted
as the marginal utility that the consumer has as decision-maker into
the electricity market. The first order approximation of $\frac{\partial G(q_{t}-\theta_{t})}{\partial q_{t}}$
when $0\leq q_{t}\leq q_{t}^{*}+\frac{p}{\gamma}$ around $q_{t}^{*}$
is

\[
\frac{\partial G(q_{t}-\theta_{t})}{q_{t}}=p-\gamma\left(q_{t}-q_{t}^{*}\right)
\]

where $-\gamma$ is the second derivative of $G(\cdotp)$ when $0\leq q_{t}\leq q_{t}^{*}+\frac{p}{\gamma}$.

\subsection{Rebate definition}

Basically, DR programs request customers to curtail demand in response
to a price signal or economic incentive. Typically the invitation
to reduce demand is made for a specific time period. There are three
main concepts: 
\begin{defn}
Baseline: The amount of energy the user would have consumed in the
absence of a request to reduce (counterfactual model) \citep{Deng2015}.
This quantity can not be measured, then it is estimated from the previous
consumption of the agent, i.e., the baseline takes into account $q_{t-1},...,q_{t-n}.$
Where $n$ defines the historical consumer behavior, i.e., $n$ corresponds
to the periods taken into account within the baseline function. 
\begin{equation}
\mathrm{Baseline}=b(q_{t-1},...,q_{t-n})\label{eq:baseline}
\end{equation}

\end{defn}

\begin{defn}
Actual Use ($q_{t}$): The amount of energy the customer actually
consumes during the event period. 
\end{defn}

\begin{figure}
\begin{centering}
\includegraphics[scale=0.8]{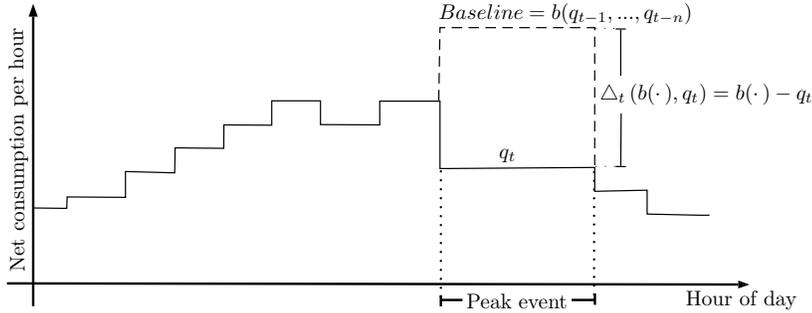}
\par\end{centering}

\caption{\label{fig:Baseline-and-rebate}Baseline and rebate definition}
\end{figure}

\begin{defn}
Load Reduction ($\triangle_{t}\left(b(\cdotp),q_{t}\right)$): The
difference between the baseline and the actual use. 

\[
b-q_{t}=\triangle_{t}\left(b(q_{t-1},...,q_{t-n}),q_{t}\right)
\]

\end{defn}
In PTR programs, the rebate is only received if there is an energy
reduction. Otherwise, the user does not get any incentive or penalty
(see fig. \ref{fig:Baseline-and-rebate}). Mathematically, 

\begin{defn}
Let $p_{2}$ the rebate price received by the user due to energy reduction
in peak periods. The PTR incentive $\pi_{2}$ is

{\scriptsize{}
\[
\pi_{2}\left(b\left(q_{t-1},...,q_{t-n}\right),q_{t}\right)=\left\{ \begin{array}{cc}
p_{2}\left(\triangle_{t}\left(b(q_{t-1}...,q_{t-n}),q_{t}\right)\right)=p_{2}(b(q_{t-1},...,q_{t-n})-q_{t}) & q_{t}<b\\
0 & q_{t}\geq b
\end{array}\right.
\]
}{\scriptsize \par}

The consumer payoff function when he is enrolled in a PTR program
is 
\begin{equation}
U_{t}\left(q_{t},\theta_{t},b(q_{t-1},...,q_{t-n})\right)=G(q_{t}-\theta_{t})-\pi(q_{t})+r\pi_{2}(b(q_{t-1},...,q_{t-n}),q_{t})\label{eq:ut}
\end{equation}

\end{defn}
where $r$ is a particular realization of a binary random variable
$R$ representing whether the consumer is called to participate in
the program according to what the SO decides.

\section{General problem formulation}

The theories of Von-Neumann and Morgenstern are employed here to model
decision-making under uncertainty. That is, the agent is assumed to
behave as if he maximizes the expected value of the payoff function according to his actions and possible consequences. The reader can found more informaction in \citep{VonNeumann1944,VegaRedondo2003,Osborne1995}. Subsequently, the consumer problem
is to find the optimal decision when he is going to participate in
a PTR program in order to increase his personal well-being and economic
profit. In addition, the optimization formulae must include all the
possible stochastic scenarios given the uncertainty of the variable
$\theta$. 

The general problem formulation from the demand side is:

{\scriptsize{}
\begin{equation}
\max_{q_{t},...,q_{t-n}\in\left[0,q_{max}\right]}\;E\left[U_{t-n}\left(q_{t-n},\theta_{t-n}\right)+...+U_{t-1}\left(q_{t-1},\theta_{t-1}\right)+U_{t}\left(q_{t},\theta_{t},b(q_{t-1},...,q_{t-n})\right)\right]\label{eq:general}
\end{equation}
}{\scriptsize \par}

where $\mathrm{\mathit{E\left[\cdot\right]}}$ is the expectation
operator. The optimization problem takes $n-1$ previous decisions
to determine the best choice for all periods including the choice
at the time $t$. Notice that the rebate price is only received at
the period $t$ (present time), namely, the payoff function at the
time $t$ is given by (\ref{eq:ut}). 

It is important to claim that the baseline could be estimated using
several techniques according to the energy policies of each country
or state. In \citep{Mohajeryami2016} some methods for baseline calculation
are found. The proposed solution for (\ref{eq:general}) is to formulate
$n$-stages optimization problems solved backward in time. At stage
$i$, the realization of $\theta_{i}$ is known. The stochastic programming
algorithm,

{\footnotesize{}$1.\,\boldsymbol{q_{t}^{o}\left(q_{t-1},...,q_{t-n};\theta_{t}\right)}=\mathrm{argmax}{}_{q_{t}\in\left[0,q_{max}\right]}\;U_{t}\left(q_{t},\theta_{t},b(q_{t-1},...,q_{t-n})\right)$}{\footnotesize \par}

{\footnotesize{}$2.\,\boldsymbol{q_{t-1}^{o}\left(q_{t-2},...,q_{t-n};\theta_{t-1}\right)}=\mathrm{argmax}{}_{q_{t-1}\in\left[0,q_{max}\right]}\;U_{t-1}\left(q_{t-1},\theta_{t-1}\right)$}{\footnotesize \par}

{\footnotesize{}$+E\left[U_{t}\left(\boldsymbol{q_{t}^{o}},\theta_{t},b(q_{t-1},...,q_{t-n})\right)\right]$}{\footnotesize \par}

$\vdots$

{\footnotesize{}$n.\,\boldsymbol{\boldsymbol{q_{t-n}^{o}(\theta_{t-n})}}=\mathrm{argmax}{}_{q_{t-n}\in\left[0,q_{max}\right]}\;U_{t-n}\left(q_{t-n},\theta_{t-n}\right)+$}{\footnotesize \par}

{\footnotesize{}$E\left[U_{t-(n-1)}\left(\boldsymbol{q_{t-n}^{o}},\theta_{t-n}\right)+...+U_{t-1}\left(\boldsymbol{q_{t-1}^{o}},\theta_{t-1}\right)+U_{t}\left(\boldsymbol{q_{t}^{o}},\theta_{t},b(\boldsymbol{q_{t-1}^{o}},\boldsymbol{q_{t-2}^{o}}...,q_{t-n})\right)\right]$}{\footnotesize \par}

It is vital to highlight that each power consumption period considered
in this algorithm has similar features of consumption, i.e., consumer
preferences and energy costs are the same in each period. For instance,
the period between 7 and 8 pm for a week. 

This paper focuses on the way to solve (\ref{eq:general}) for finding
a closed form result for the consumer decision. In the next section,
The stochastic optimization problem is solved for two periods. Furthermore,
it is assumed that the user is always called to participate in PTR
program, henceforth $r=1$ is considered.

\section{Problem formulation for two periods}

A single previous period $t-1$ is assumed to estimate the baseline
in eq. (\ref{eq:baseline}). Then, the baseline is $b(q_{t-1})=q_{t-1}$.
In this regard, the problem formulation is:

\begin{equation}
\max_{q_{t},q_{t-1}\in\left[0,q_{max}\right]}\;E\left[U_{t-1}\left(q_{t-1},\theta_{t-1}\right)+U_{t}\left(q_{t},\theta_{t},b(q_{t-1})\right)\right]\label{eq:problem}
\end{equation}

First, the agent maximizes the energy consumption at the \textquotedbl{}present\textquotedbl{}
time $t$, given that the realization of $\theta_{t}$ and the value
of $q_{t-1}$ are known.

\[
\boldsymbol{q_{t}^{o}\left(q_{t-1};\theta_{t}\right)}=\mathrm{arg\mathrm{max}}_{q_{t}\in\left[0,q_{max}\right]}\;U_{t}\left(q_{t},\theta_{t},b(q_{t-1})\right)
\]

Second, the decision-maker determines the best consumption for the
baseline setting period, knowing the rational choice $q_{t}^{o}$
for the future. The realization of $\theta_{t-1}$ is given and the
user faces uncertainty in $\theta_{t}$ only, i.e.,

\[
\boldsymbol{q_{t-1}^{o}(\theta_{t-1})}=\mathrm{argmax}_{q_{t-1}\in\left[0,q_{max}\right]}\;U_{t-1}\left(q_{t-1},\theta_{t-1}\right)+E\left[U_{t}\left(\boldsymbol{q_{t}^{o}},\theta_{t},b(q_{t-1})\right)\right]
\]

\subsection{First-stage stochastic programming}

The following result presents the solution $q_{t}^{o}$ to the first-stage
stochastic optimization at the time $t$. 
\begin{thm}
\label{thm:The-optimal-consumption qt}The optimal consumption $\boldsymbol{q_{t}^{o}}$
of a user participating in a PTR program (i.e. the solution of the
first-stage stochastic programming), given $G(\cdotp)$ in (\ref{eq:g})
and $U_{t}(\cdotp)$ in (\ref{eq:ut}), is:

{\small{}
\[
\boldsymbol{q_{t}^{o}\left(q_{t-1};\theta_{t}\right)}=\left\{ \begin{array}{ccc}
q_{t}^{*} & \:r=1\;\mathrm{and}\;q_{t-1}-\overline{q}+\frac{p_{2}}{2\gamma}<\theta_{t}\leq\overline{\theta} & strategy\:A\\
q_{t}^{*}-\frac{p_{2}}{\gamma} & \:r=1\;\mathrm{and}\;\frac{p_{2}}{\gamma}-\overline{q}<\theta_{t}\leq q_{t-1}-\overline{q}+\frac{p_{2}}{2\gamma} & strategy\:B\\
0 & \:r=1\:\mathrm{and}\:\underline{\theta}\leq\theta_{t}\leq\frac{p_{2}}{\gamma}-\overline{q} & strategy\:C\\
q_{t}^{*} & \;r=0 & strategy\:D
\end{array}\right.
\]
}{\small \par}
\end{thm}
Note that when the SO calls the user ($r=1$), strategy $A$ means
that the user decides rationally to spend $q^{*}$ of energy (That
is, he does not reduce energy consumption), strategy $B$ depicts
whether the consumer chooses to diminish the demand to $q_{t}^{*}-\frac{p_{2}}{\gamma}$
and finally, strategy $C$ is when the best decision is to consume zero
energy. 

\begin{figure}[H]
\begin{centering}
\includegraphics[scale=0.8]{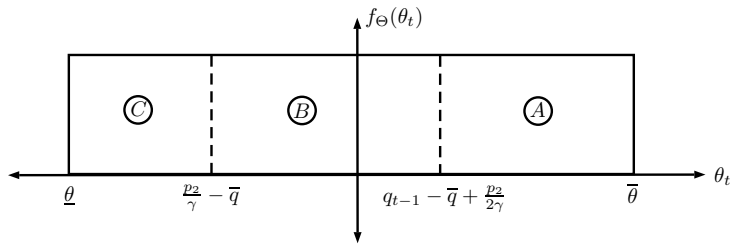} 
\par\end{centering}

\protect\protect\caption{\label{fig:Probability-density-function}User optimal strategies within
$\theta_{t}$ probability density function.}
\end{figure}

According to theorem \ref{thm:The-optimal-consumption qt}, note that
the best decision depends on the realization of $\theta_{t}$. Then,
the user chooses a strategy at the time $t$ given his actual demand.
The proof is shown in appendix A. 
\begin{cor}
\label{cor:qt}The expected value of the load $q_{t}^{o}$ is:

{\small{}
\[
E\left[\boldsymbol{q_{t}^{o}\left(q_{t-1};\theta_{t}\right)}\right]=\left\{ \begin{array}{ccc}
\overline{q} & \:r=1\;\mathrm{and}\;p_{2}\leq2\gamma\left(\overline{q}-q_{t-1}\right) & strategy\:A\\
\overline{q}-\frac{p_{2}}{\gamma} & \:r=1\;\mathrm{and}\;2\gamma\left(\overline{q}-q_{t-1}\right)<p_{2}\leq\overline{q}\gamma & strategy\:B\\
0 & \:r=1\:\mathrm{and}\:\overline{q}\gamma<p_{2} & strategy\:C\\
\overline{q} & \;r=0 & strategy\:D
\end{array}\right.
\]
}{\small \par}
\end{cor}
The expected value of consumer payoff $E\left[U_{t}\left(\boldsymbol{q_{t}^{o}},\theta_{t},b(q_{t-1})\right)\right]$
is found assuming a continuos uniform distribution function $f_{\Theta}(\theta_{t})$.
Since $r=1$, from theorem \ref{thm:The-optimal-consumption qt},
the consumer has three availables strategies according to the realization
of uncertainty $\theta_{t}$. In addition, the random variable is
symmetric with respect to zero. Whether strategies $A$, $B$ and
$C$ are feasible according to the parameters $\overline{\theta}$,
$\underline{\theta}$, $p_{2}$, $\gamma$, $\overline{q}$ and the
variable $q_{t-1}$ then these stategies is within the probability
density function of $\theta_{t}$ which it is shown in fig. \ref{fig:Probability-density-function}. 

Looking in detail the intervals of $\theta_{t}$ that define strategy
$C$, these depend of constant values, whereas the intervals for strategies
$A$ and $B$ depend on the optimization variable $q_{t-1}$. Therefore,
the probabilistic events change with $q_{t-1}$. For instance, strategy
$A$ has zero probability when $q_{t-1}>\overline{\theta}+\overline{q}-\frac{p_{2}}{2\gamma}$.

\begin{cor}
\label{cor:The-expected-value}The expected value of the payoff in
$t$, $E\left[U_{t}\left(\boldsymbol{q_{t}^{o}},\theta_{t},b(q_{t-1})\right)\right]$,
depends on the probabilities of available strategies. Therefore, a
construction by cases is employed to solve \textup{$E\left[U_{t}\left(\boldsymbol{q_{t}^{o}},\theta_{t},b(q_{t-1})\right)\right]$}.
There are three main cases:

$case\;1$: $\underline{\theta}>\frac{p_{2}}{\gamma}-\overline{q}$,
strategy $C$ does not exist. Then, $E\left[U_{t}\left(\boldsymbol{q_{t}^{o}},\theta_{t},b(q_{t-1})\right)\right]$
depends on the value of $q_{t-1}$. Therefore, 

$E\left[U_{t}\left(\boldsymbol{q_{t}^{o}},\theta_{t},b(q_{t-1})\right);\underline{\theta}>\frac{p_{2}}{\gamma}-\overline{q}\right]=$

$\left\{ \begin{array}{cc}
E_{A}=\int_{\underline{\theta}}^{\overline{\theta}}U_{t}\left(q_{t}^{*}\right)f_{\Theta}(\theta_{t})d\theta_{t} & q_{t-1}\in\left[0,\overline{q}+\underline{\theta}-\frac{p_{2}}{2\gamma}\right]\\
\begin{array}{c}
E_{AB}=\int_{q_{t-1}-\overline{q}+\frac{p_{2}}{2\gamma}}^{\overline{\theta}}U_{t}\left(q_{t}^{*}\right)f_{\Theta}(\theta_{t})d\theta_{t}+\\
\int_{\underline{\theta}}^{q_{t-1}-\overline{q}+\frac{p_{2}}{2\gamma}}\mathit{U_{t}\left(q_{t}^{*}-\frac{p_{2}}{\gamma}\right)}f_{\Theta}(\theta_{t})d\theta_{t}
\end{array} & \:q_{t-1}\in\left[\overline{q}+\underline{\theta}-\frac{p_{2}}{2\gamma},\overline{q}+\overline{\theta}-\frac{p_{2}}{2\gamma}\right]\\
E_{B}=\int_{\underline{\theta}}^{\overline{\theta}}U_{t}\left(q_{t}^{*}-\frac{p_{2}}{\gamma}\right)f_{\Theta}(\theta_{t})d\theta_{t} & q_{t-1}\in\left[\overline{q}+\overline{\theta}-\frac{p_{2}}{2\gamma},q_{max}\right]
\end{array}\right.$

$case\;2$: $\underline{\theta}\leq\frac{p_{2}}{\gamma}-\overline{q}<\overline{\theta}$,
strategy $C$ has positive probability. Therefore, $E\left[U_{t}\left(\boldsymbol{q_{t}^{o}},\theta_{t},b(q_{t-1})\right)\right]$
is given by:

$E\left[U_{t}\left(\boldsymbol{q_{t}^{o}},\theta_{t},b(q_{t-1})\right);\underline{\theta}\leq\frac{p_{2}}{\gamma}-\overline{q}<\overline{\theta}\right]=$

$\left\{ \begin{array}{cc}
E_{A}=\int_{\underline{\theta}}^{\overline{\theta}}U_{t}\left(q_{t}^{*}\right)f_{\Theta}(\theta_{t})d\theta_{t} & q_{t-1}\in\left[0,\overline{q}+\underline{\theta}-\frac{p_{2}}{2\gamma}\right]\\
\begin{array}{c}
E_{AC}=\int_{q_{t-1}-\overline{q}+\frac{p_{2}}{2\gamma}}^{\overline{\theta}}U_{t}\left(q_{t}^{*}\right)f_{\Theta}(\theta_{t})d\theta_{t}+\\
\int_{\underline{\theta}}^{q_{t-1}-\overline{q}+\frac{p_{2}}{2\gamma}}\mathit{U_{t}\left(0\right)}f_{\Theta}(\theta_{t})d\theta_{t}
\end{array} & q_{t-1}\in\left[\overline{q}+\underline{\theta}-\frac{p_{2}}{2\gamma},\frac{p_{2}}{2\gamma}\right]\\
\begin{array}{c}
E_{ABC}=\int_{q_{t-1}-\overline{q}+\frac{p_{2}}{2\gamma}}^{\overline{\theta}}U_{t}\left(q_{t}^{*}\right)f_{\Theta}(\theta_{t})d\theta_{t}+\\
\int_{\frac{p_{2}}{\gamma}-\overline{q}}^{q_{t-1}-\overline{q}+\frac{p_{2}}{2\gamma}}\mathit{U_{t}\left(q_{t}^{*}-\frac{p_{2}}{\gamma}\right)}f_{\Theta}(\theta_{t})d\theta_{t}+\\
\int_{\underline{\theta}}^{\frac{p_{2}}{\gamma}-\overline{q}}U_{t}\left(0\right)f_{\Theta}(\theta_{t})d\theta_{t}
\end{array} & q_{t-1}\in\left[\frac{p_{2}}{2\gamma},\overline{q}+\overline{\theta}-\frac{p_{2}}{2\gamma}\right]\\
\begin{array}{c}
E_{BC}=\int_{\frac{p_{2}}{\gamma}-\overline{q}}^{\overline{\theta}}U_{t}\left(q_{t}^{*}-\frac{p_{2}}{\gamma}\right)f_{\Theta}(\theta_{t})d\theta_{t}+\\
\int_{\underline{\theta}}^{\frac{p_{2}}{\gamma}-\overline{q}}U_{t}\left(0\right)f_{\Theta}(\theta_{t})d\theta_{t}
\end{array} & q_{t-1}\in\left[\overline{q}+\overline{\theta}-\frac{p_{2}}{2\gamma},q_{max}\right]
\end{array}\right.$

$case\;3$: $\frac{p_{2}}{\gamma}-\overline{q}\geq\overline{\theta}$,
a priori, strategy $C$ has probability one. However, the main point
is $q_{t-1}-\overline{q}+\frac{p_{2}}{\gamma}$ then it could be exist
other strategies different from $C$. Thus, $E\left[U_{t}\left(\boldsymbol{q_{t}^{o}},\theta_{t},b(q_{t-1})\right)\right]$
is given by:

$E\left[U_{t}\left(\boldsymbol{q_{t}^{o}},\theta_{t},b(q_{t-1})\right);\frac{p_{2}}{\gamma}-\overline{q}\geq\overline{\theta}\right]=$

$\left\{ \begin{array}{cc}
E_{A}=\int_{\underline{\theta}}^{\overline{\theta}}U_{t}\left(q_{t}^{*}\right)f_{\Theta}(\theta_{t})d\theta_{t} & q_{t-1}\in\left[0,\overline{q}+\underline{\theta}-\frac{p_{2}}{2\gamma}\right]\\
\begin{array}{c}
E_{AC'}=\int_{q_{t-1}-\overline{q}+\frac{p_{2}}{2\gamma}}^{\overline{\theta}}U_{t}\left(q_{t}^{*}\right)f_{\Theta}(\theta_{t})d\theta_{t}+\\
\int_{\underline{\theta}}^{q_{t-1}-\overline{q}+\frac{p_{2}}{2\gamma}}\mathit{U_{t}\left(0\right)}f_{\Theta}(\theta_{t})d\theta_{t}
\end{array} & q_{t-1}\in\left[\overline{q}+\underline{\theta}-\frac{p_{2}}{2\gamma},\overline{q}+\overline{\theta}-\frac{p_{2}}{2\gamma}\right]\\
E_{C}==\int_{\underline{\theta}}^{\overline{\theta}}U_{t}\left(0\right)f_{\Theta}(\theta_{t})d\theta_{t} & q_{t-1}\in\left[\overline{q}+\overline{\theta}-\frac{p_{2}}{2\gamma},q_{max}\right]
\end{array}\right.$
\end{cor}
Note that the expected value $E\left[\mathit{U_{t}\left(\boldsymbol{q_{t}^{o}},\theta_{t},q_{t-1}\right)}\right]$
is a piecewise function that depends on the value of $q_{t-1}$.

\subsection{Second-stage stochastic programming }

For the second-stage, the rational choice for $\boldsymbol{q_{t}^{o}}$
is known and the realization of $\theta_{t-1}$ is given. Then $\boldsymbol{q_{t-1}^{o}}$
is found by using the result of theorem \ref{thm:The-optimal-consumption qt}.
The optimization problem is:

{\footnotesize{}
\begin{equation}
\boldsymbol{q_{t-1}^{o}(\theta_{t-1})}=\mathrm{argmax}_{q_{t-1}\geq0}\;G\left(q_{t-1}-\theta_{t-1}\right)-pq_{t-1}+E\left[G\left(\boldsymbol{q_{t}^{o}}-\theta_{t}\right)-p\boldsymbol{q_{t}^{o}}+r\pi_{2}\left(q_{t-1},\boldsymbol{q_{t}^{o}}\right)\right]\label{eq:second_stage}
\end{equation}
}{\footnotesize \par}

The mathematical solution of (\ref{eq:second_stage}) is developed
in the following three theorems for each case mentioned in the corollary
\ref{cor:The-expected-value} and the proofs are found in appendixes
B, C, and D.
\begin{thm}
\label{th case 1}Given $\underline{\theta}>\frac{p_{2}}{\gamma}-\overline{q}$
(case 1) and $\overline{q}+\frac{p}{\gamma}>\overline{q}+\overline{\theta}-\frac{p_{2}}{2\gamma}$,
then the optimal solution $\boldsymbol{q_{t-1}^{o}}$ for (\ref{eq:second_stage})
is:

\[
E\left[\boldsymbol{q_{t-1}^{o}(\theta_{t-1})}\right]=\left\{ \begin{array}{cc}
\overline{q}-\frac{p_{2}}{2\gamma}+\frac{2p_{2}\overline{\theta}}{2\overline{\theta}\gamma-p_{2}} & 0\leq p_{2}<\frac{2}{3}\overline{\theta}\gamma\\
\overline{q}+\frac{p_{2}}{\gamma} & \frac{2}{3}\overline{\theta}\gamma\leq p_{2}<p\\
q_{max} & p\leq p_{2}<\gamma\left(\underline{\theta}+\overline{q}\right)
\end{array}\right.
\]

\end{thm}

\begin{thm}
\label{th case 2}Given $\underline{\theta}\leq\frac{p_{2}}{\gamma}-\overline{q}<\overline{\theta}$
(case 2) and $\overline{q}+\frac{p}{\gamma}>\overline{q}+\overline{\theta}-\frac{p_{2}}{2\gamma}$,
then the optimal solution $\boldsymbol{q_{t-1}^{o}}$ for (\ref{eq:second_stage})
is:

\[
E\left[\boldsymbol{q_{t-1}^{o}(\theta_{t-1})}\right]=\left\{ \begin{array}{cc}
\overline{q}-\frac{p_{2}}{2\gamma}+\frac{2p_{2}\overline{\theta}}{2\overline{\theta}\gamma-p_{2}} & \gamma\left(\underline{\theta}+\overline{q}\right)\leq p_{2}<\frac{2}{3}\overline{\theta}\gamma\\
\overline{q}+\frac{p_{2}}{\gamma} & \frac{2}{3}\overline{\theta}\gamma\leq p_{2}<p\\
q_{max} & p<p_{2}\leq\gamma\left(\overline{\theta}+\overline{q}\right)
\end{array}\right.
\]

\end{thm}

\begin{thm}
\label{th case 3}Given $\frac{p_{2}}{\gamma}-\overline{q}\geq\overline{\theta}$
(case 3) and $\overline{q}+\frac{p}{\gamma}>\overline{q}+\overline{\theta}-\frac{p_{2}}{2\gamma}$,
then the optimal solution $\boldsymbol{q_{t-1}^{o}}$ for (\ref{eq:second_stage})
is:

\[
E\left[\boldsymbol{q_{t-1}^{o}(\theta_{t-1})}\right]=\left\{ \begin{array}{cc}
\overline{q}+\frac{p_{2}}{\gamma} & \gamma\left(\overline{\theta}+\overline{q}\right)\leq p_{2}<p\\
q_{max} & p_{2}>p
\end{array}\right.
\]

\end{thm}
Theorems \ref{th case 1}, \ref{th case 2} and \ref{th case 3} present
the optimal consumption $q_{t-1}$ given the solutions of theorem
\ref{thm:The-optimal-consumption qt}. For theorem \ref{th case 1},
the result is rigthful for incentives less than $\gamma\left(\underline{\theta}+\overline{q}\right)$,
which means that strategy $C$ does not exist. In addition, the saturation
part of the consumer (see equation (\ref{eq:g})) is $\overline{q}+\overline{\theta}-\frac{p_{2}}{2\gamma}<\overline{q}+\frac{p}{\gamma}<q_{max}$,
namely, when $E\left[U_{t}\left(\boldsymbol{q_{t}^{o}},\theta_{t},b(q_{t-1})\right);\underline{\theta}>\frac{p_{2}}{\gamma}-\overline{q}\right]$
is strategy $B$, specifically, $q_{t-1}\in\left[\overline{q}+\overline{\theta}-\frac{p_{2}}{2\gamma},q_{max}\right]$.
Whether the user has low uncertainty, the theorem \ref{th case 1}
is employed for estimating optimal decision at the time $t-1$. Note
that if $0\leq p_{2}<\frac{2}{3}\overline{\theta}\gamma$ the solutions
is decreasing with respect to $p_{2}$, therefore, the situation when
the incentive is too small, it is risky to increase the energy consumption
at the baseline setting period. Nonetheless, this event is not common
owing to the incentive is equal or greater than retail price. Next,
whether $\frac{2}{3}\overline{\theta}\gamma\leq p_{2}<p$ then the
optimal strategies is to increase $\overline{q}+\frac{p_{2}}{\gamma}$.
Finally, if the incentive is greater than $p$ then the optimal choice
is to increase the energy consumption as much as possible. Moreover,
The meaning of theorem \ref{th case 2} is the same than the theorem
\ref{th case 2}. However, the consumer uncertainty is larger and
the incentive limit is given by $\gamma\left(\underline{\theta}+\overline{q}\right)<p_{2}<\gamma\left(\overline{\theta}+\overline{q}\right)$.
Finally, theorem \ref{th case 3} is valid for $p_{2}\geq\gamma\left(\overline{\theta}+\overline{q}\right)$
and $p>\gamma\overline{\theta}-\frac{p_{2}}{2}$. Note that there
are only two solutions that depend on incentive $p_{2}$. The uncertainty
is greater than the previous two theorems. In general, The saturation
of consumer preferences causes that the user wastes energy.

\section{Numerical examples}

In this section, simulation results are presented to illustrate the
optimal behavior of a user when he is participating in a PTR program.
The utility function for this example is

\[
G\left(q_{t}-\theta_{t}\right)=\left\{ \begin{array}{cc}
-\frac{\gamma}{2}\left(q_{t}-q_{t}^{*}\right)^{2}+p\left(q_{t}-q_{t}^{*}\right)+\frac{\gamma}{2}\overline{q}^{2}+p\overline{q} & \quad0\leq q_{t}\leq\overline{q}+\frac{p}{\gamma}+\theta_{t}\\
p\overline{q}+\frac{p^{2}}{2\gamma}+\frac{\gamma\overline{q}^{2}}{2} & \quad q_{t}>\overline{q}+\frac{p}{\gamma}+\theta_{t}
\end{array}\right.
\]

The retail price is $p= 0.26\$/kWh$ (based on peak summer rate in 10/1/16 by Pacific Gas and Electric Company in San Francisco, California), deterministic baseline $\overline{q}=8kWh$
and the curvature $\gamma=0.05$. Randomness $\theta_{t}$ for each
period has been created as a uniform random variable with zero mean
and with simetric support. A Monte Carlo simulation is performed with
10000 realizations of $\theta_{t}$ for each value of $q_{t-1}$. 

\begin{figure}[H]
\begin{centering}
\includegraphics[scale=0.55]{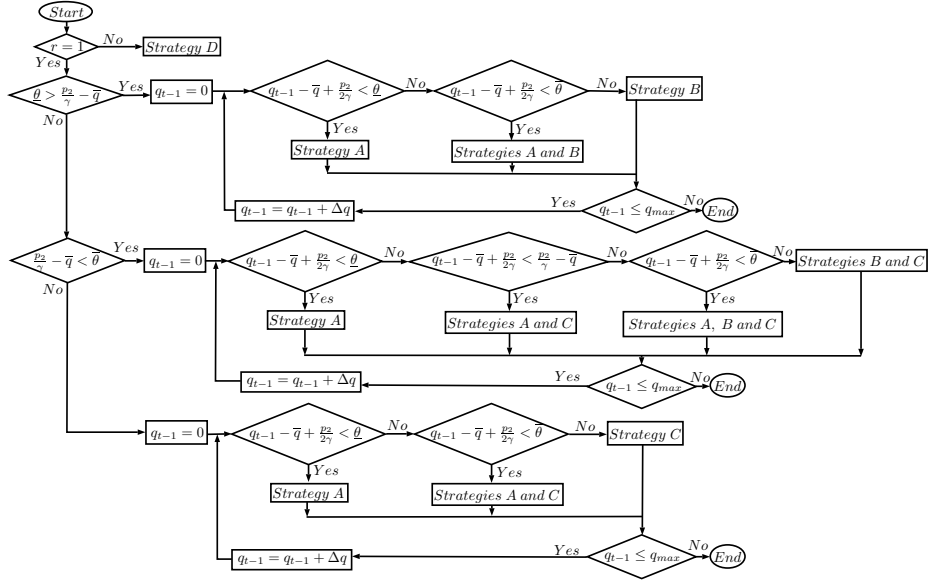} 
\par\end{centering}

\caption{\label{fig:Flowchart}Flowchart to calculate the expected value of
$U_{t}(\cdotp)$}
\end{figure}

The flowchart in fig. \ref{fig:Flowchart}. shows the conditions to
determine the case that the user faces for choosing his decision according
to the value of $q_{t-1}$, in order to solve the expected value in
(\ref{eq:second_stage}). This flowchart is derived from fig. \ref{fig:Probability-density-function}
and corollary \ref{cor:The-expected-value} by analyzing when strategies
have positive probability.

\subsection{Incentives analysis}

\begin{figure}
\begin{centering}
\begin{tabular}{|c|c|c|}
\hline 
$p_{2}$ & $q_{t}$ & $E\left[U_{t-1}+U_{t}\right]$\tabularnewline
\hline 
\hline 
\multirow{1}{*}{$0.15$} & \includegraphics[scale=0.35]{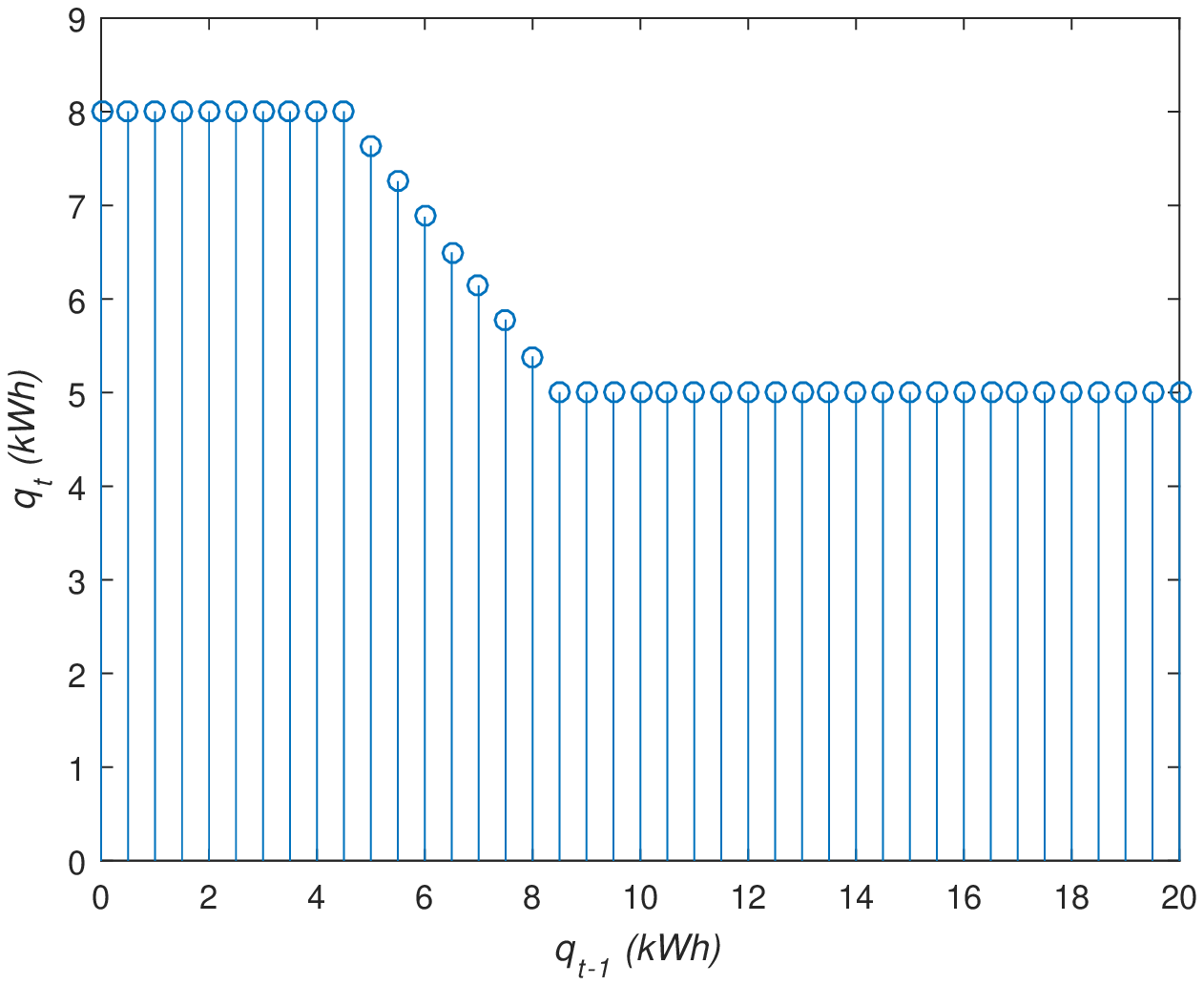} & \includegraphics[scale=0.35]{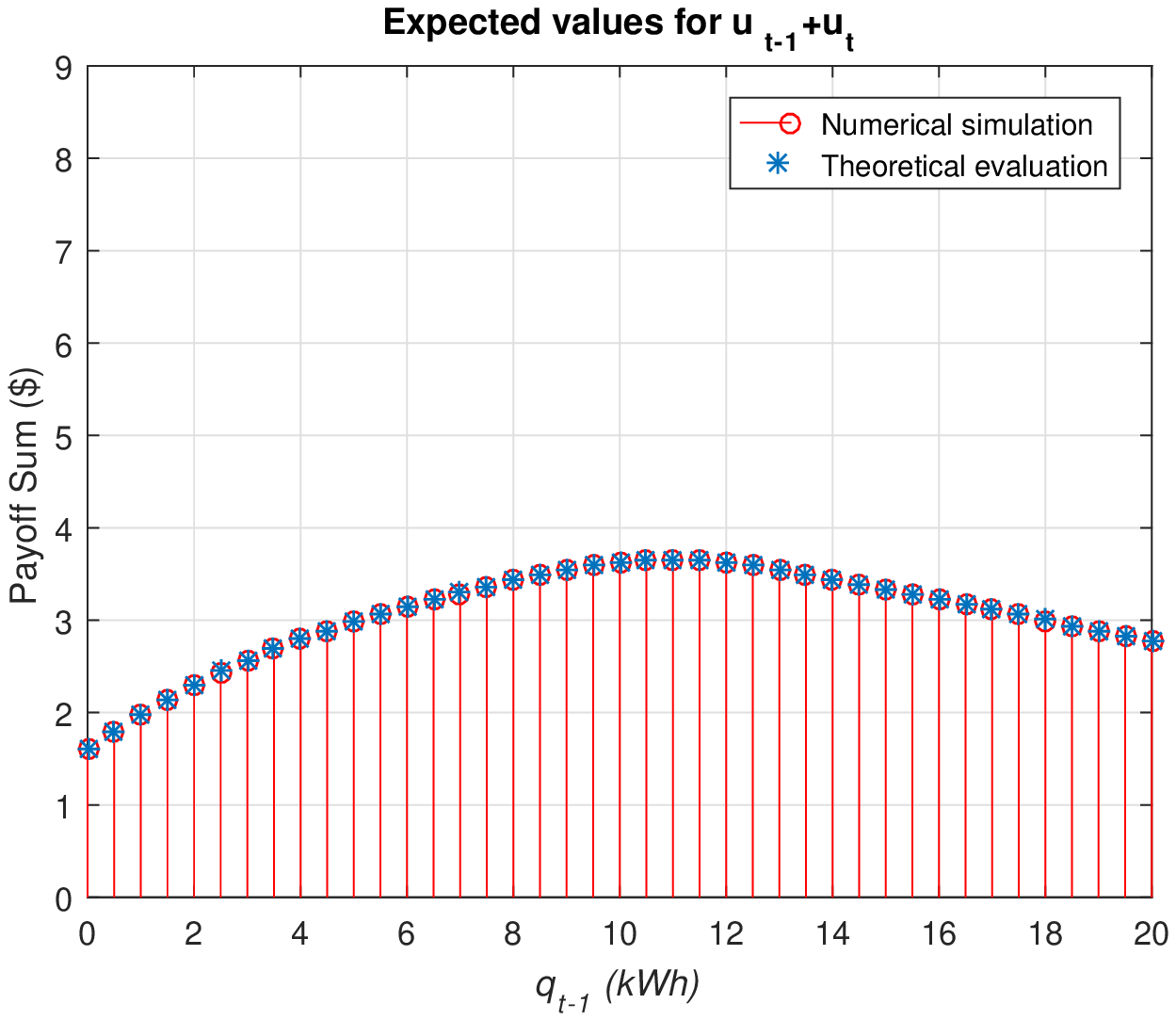}\tabularnewline
\hline 
$0.26$ & \includegraphics[scale=0.35]{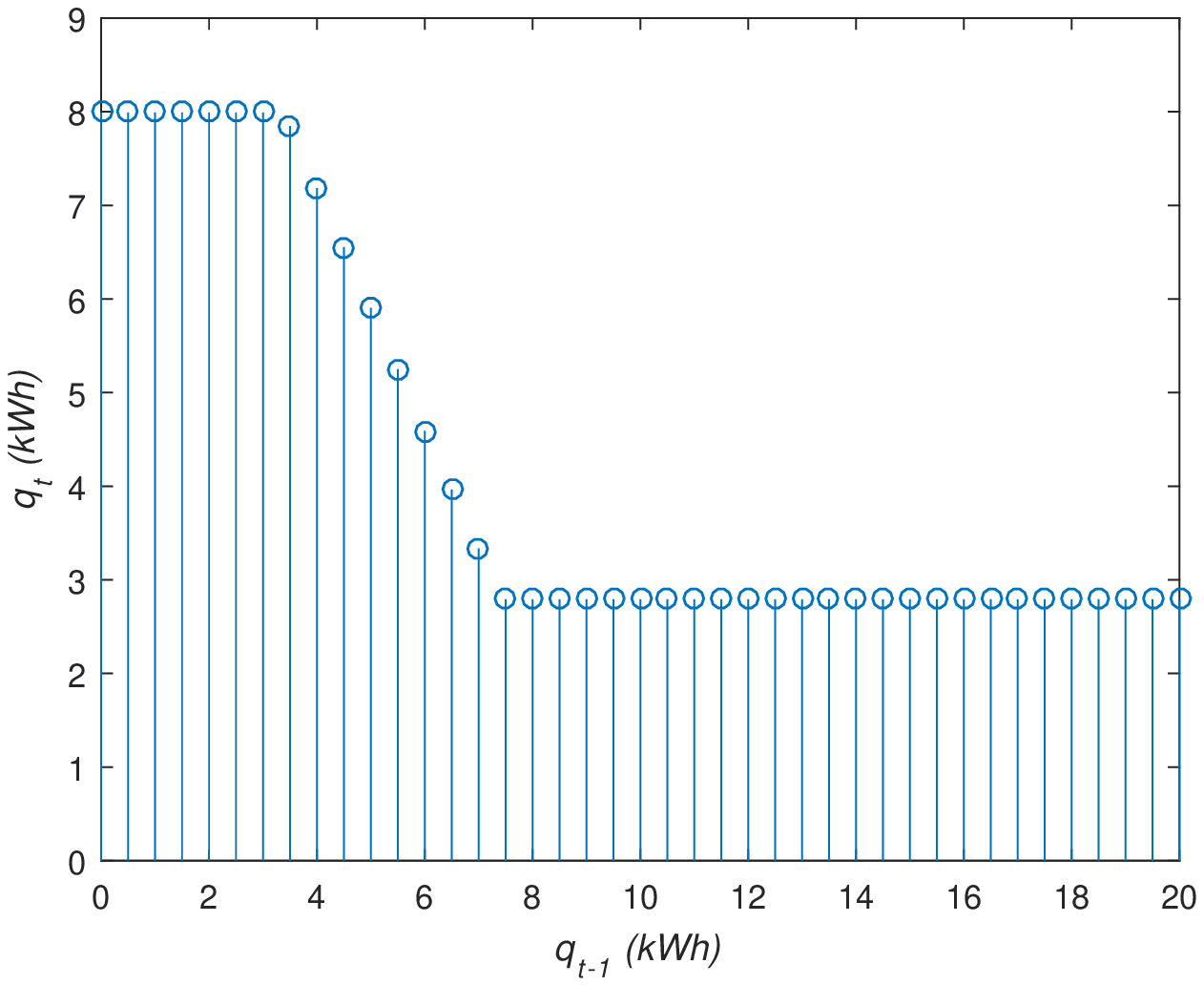} & \includegraphics[scale=0.35]{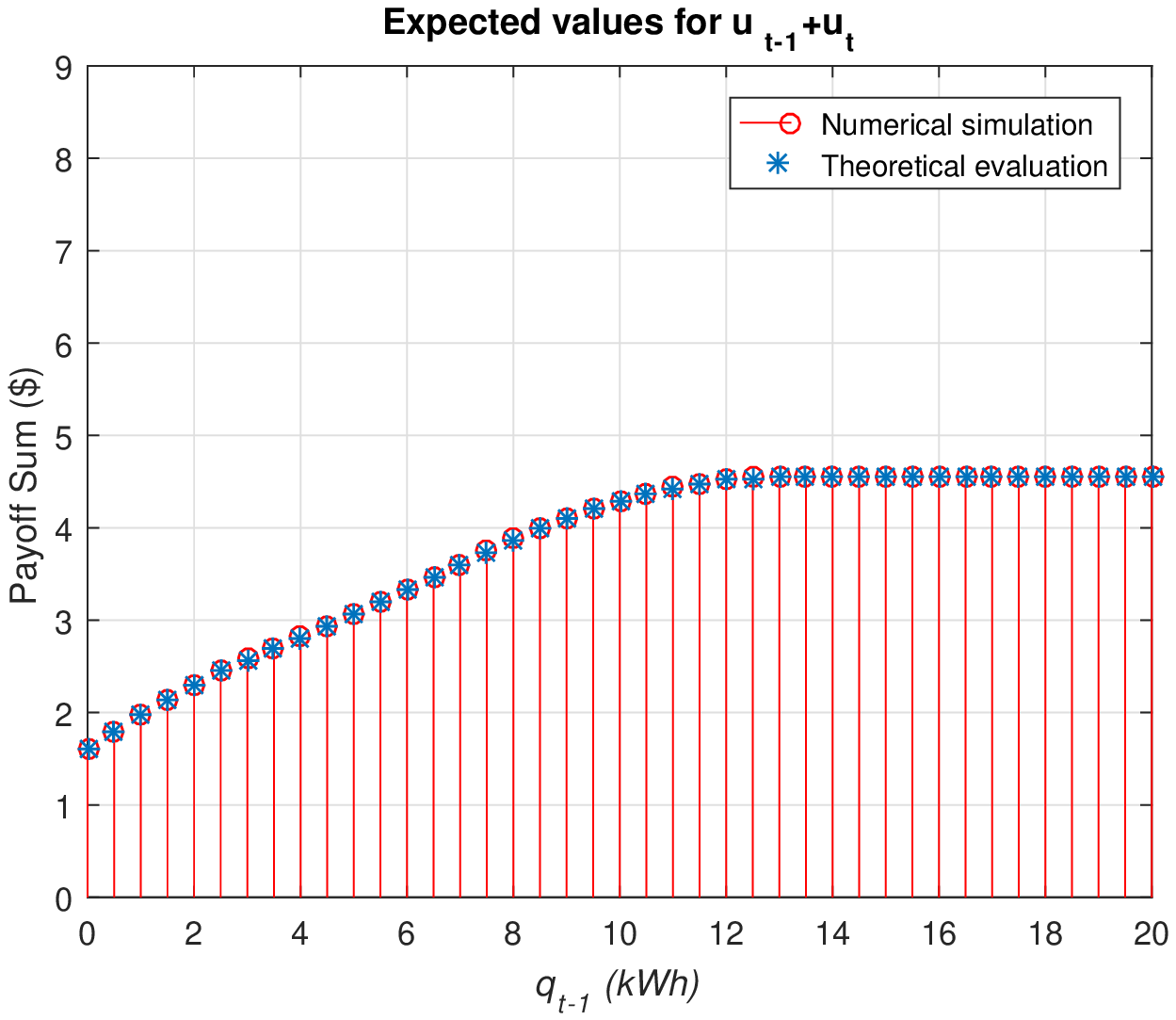}\tabularnewline
\hline 
$0.45$ & \includegraphics[scale=0.35]{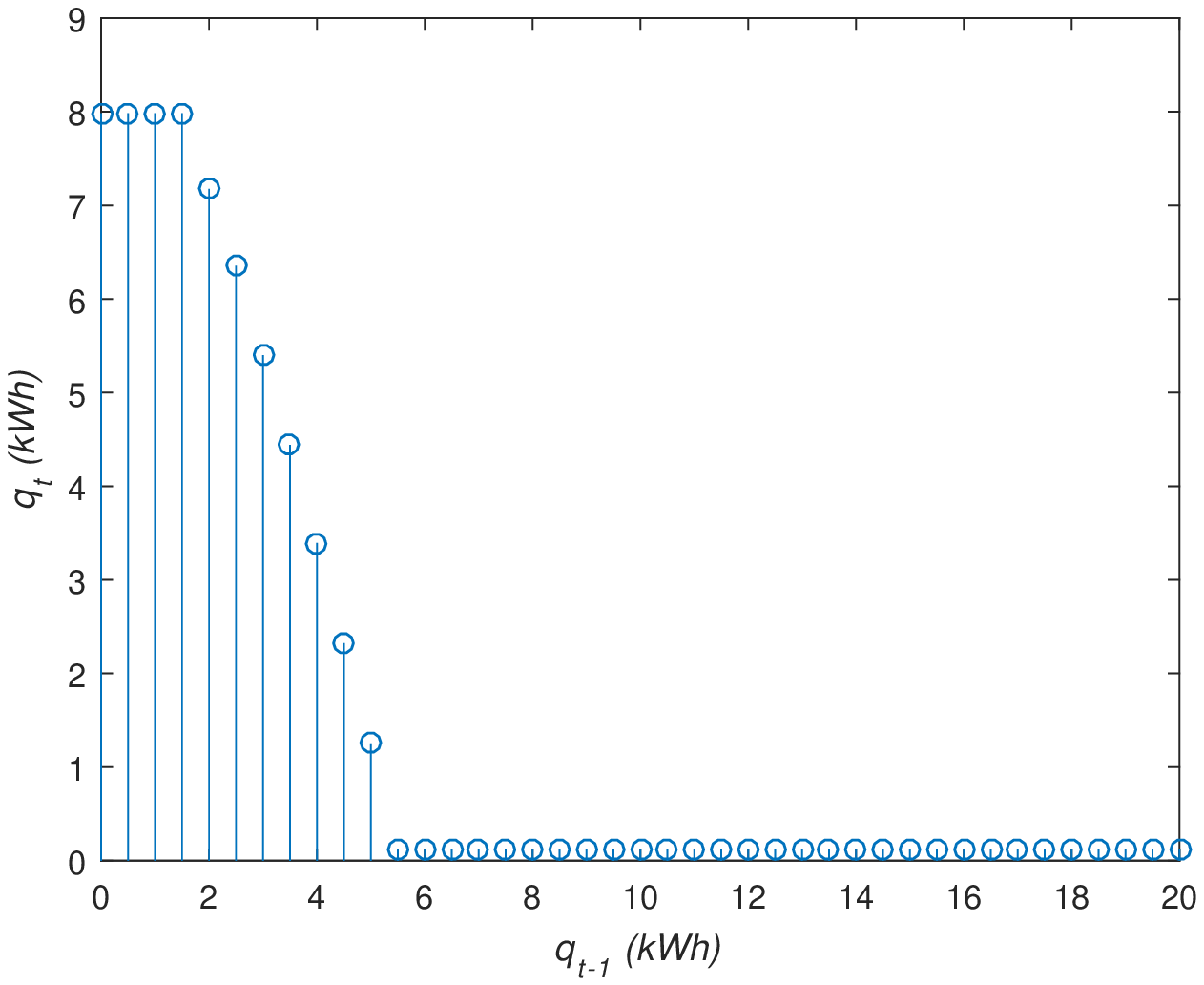} & \includegraphics[scale=0.35]{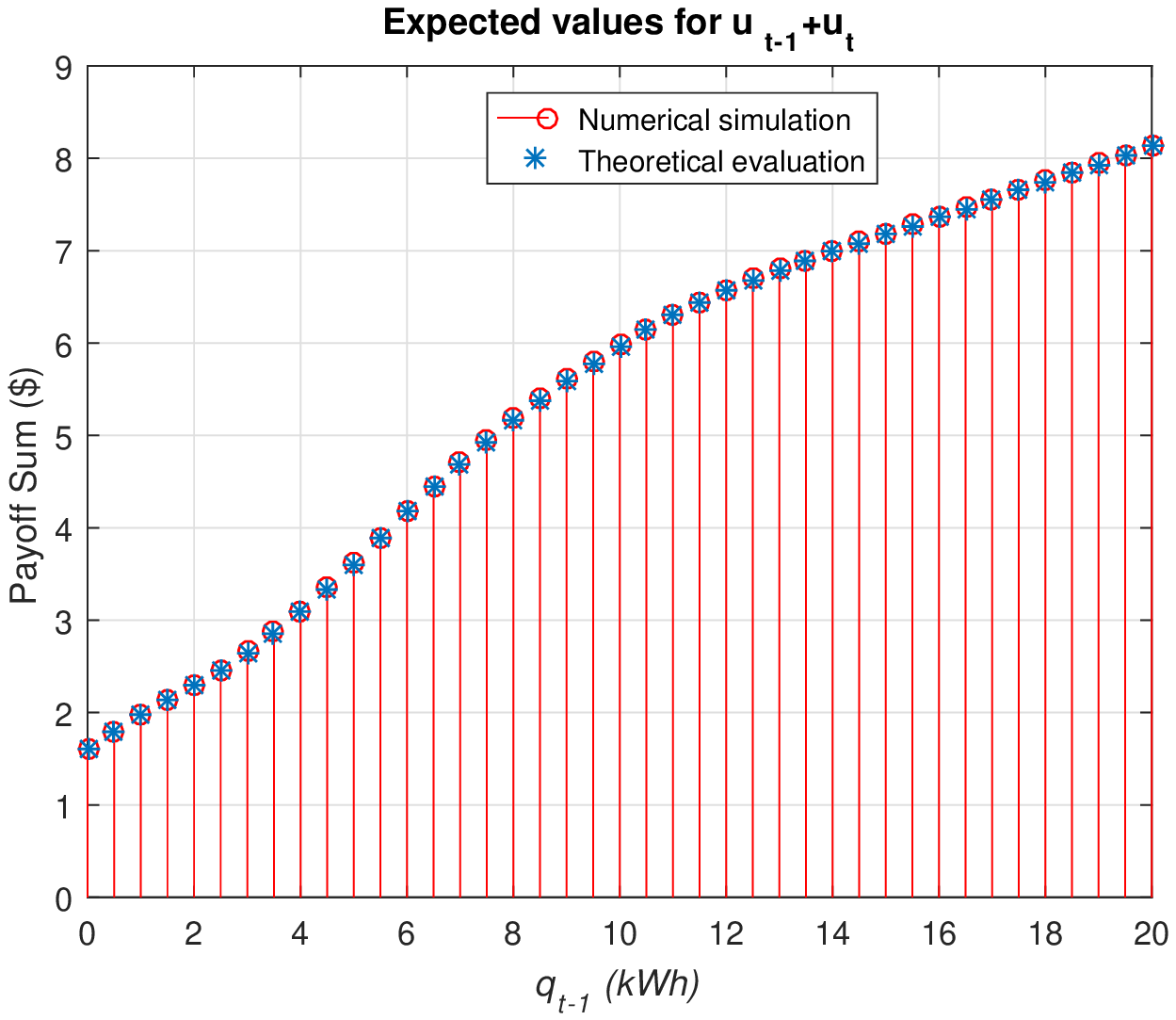}\tabularnewline
\hline 
\end{tabular}
\par\end{centering}

\caption{\label{fig:Incentives-analysis}Incentives analysis.}
\end{figure}

In this subsection, the effect of the incentive $p_{2}$ on the load
and user utility at time $t$ given the baseline $q_{t-1}$ is studied
by changing the reward $p_{2}$. For this analysis, $\theta_{t}\sim\mathrm{unif}\left[-0.25\overline{q},0.25\overline{q}\right]$
and $q_{max}=20kWh$ are assumed. In fig. \ref{fig:Incentives-analysis}
are shown three different situations that depend on the incentive
value. The first column, the reward $p_{2}$ is presented for each
situation. The second one, the plot of energy consumptions at the
time $t$ versus the consumption at the time $t-1$ are shown according
to the incentive. Third column, the expected value of profit function
based on the decisions at time $t-1$.

First, the event when the incentive is lower than retail price, i.e.,
$p_{2}<p$ is evaluated. For $p_{2}= 0.15\$/kWh$, the optimal solution
is to increase energy consumption at the period $t-1$, close to $q_{t-1}=\overline{q}+\theta_{t-1}+\frac{p_{2}}{\gamma}=11kWh$
and reduce energy consumption at $t$ to $q_{t}^{o}=\overline{q}+\theta_{t}-\frac{p_{2}}{\gamma}=5kWh$.
Next, when the incentive and the retail price are the same $p_{2}=p$.
The rational user consumes at the past time whatever value comprising
in $q_{t-1}=[11kWh,\:20kWh]$, then, taking into account the worst
event, the user consumes $q_{t-1}=20kWh$, therefore, the optimal consumption
at the time $t$, it is $q_{t}^{o}=\overline{q}+\theta_{t}-\frac{p_{2}}{\gamma}=2.79kWh$.
Finally, the situation when the incentive is greater than retail price,
namely, $p_{2}>p$ is assessed. For $p_{2}=0.45\$/kWh$, the optimal
behavior is to consume as much energy as possible, $q_{t-1}=q_{max}$,
irrespective of parameters $\gamma$, $\overline{q}$, $\overline{\theta}$
and $\underline{\theta}$. If the maximum value is $q_{t-1}=20kWh$
then, he would consume zero energy $q_{t}=0kWh$ at the time $t$
in order to get the maximum profit. These behaviors are predicted
by corollary \ref{cor:qt} and theorem \ref{th case 1}.

\begin{table}[H]
\begin{centering}
\begin{tabular}{|c|c|c|c|c|}
\hline 
$p_{2}$ ($\$/kWh$)  & Expected profit (\$)  & $q_{t-1}^{o}$ ($kWh$)  & $q_{t}^{o}$ ($kWh$)  & $q_{t-1}^{o}+q_{t}^{o}$ ($kWh$)\tabularnewline
\hline 
\hline 
0 & 3.2 & 8 & 8 & 16\tabularnewline
\hline 
0.15  & 3.65  & 11  & 5  & 16\tabularnewline
\hline 
0.26 & 4.55 & 20 & 2.79 & 22.79\tabularnewline
\hline 
0.45  & 8.13 & 20  & 0  & 20\tabularnewline
\hline 
\end{tabular}
\par\end{centering}

\caption{\label{tab:Profit-comparison}Comparison of optimal user strategies
under different incentives.}
\end{table}

\begin{figure}[H]
\begin{centering}
\includegraphics[scale=0.7]{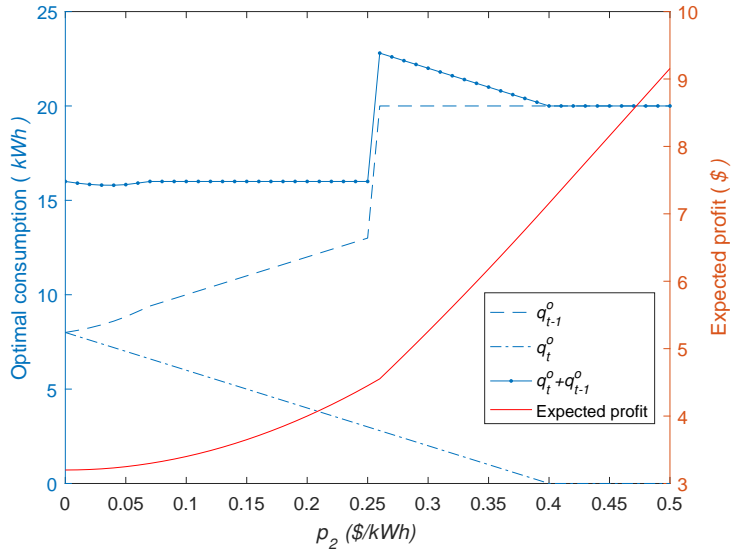}
\par\end{centering}

\caption{\label{fig:incentive summary}Optimal consumption and profit with
25\% of uncertainty.}
\end{figure}

The previous results are summarized in table \ref{tab:Profit-comparison}.
Whether the user is not called or the incentive is zero then the trivial
solution is not to alter his behavior. On the other hand, when the
price incentive is higher than zero but lower than the retail price
($p_{2}=0.15$), the user is induced to raise his consumption to alter
the baseline and get the highest economic benefits by reducing the
consumption at time $t$, getting a profit of \$3.65. Likewise, whether
the agent gets an incentive equal or greater than the retail price
then he alters his consumption up to the maximum possible load to
maximize the profit to \$4.55 or \$8.13 according to the incentive, consuming
much more energy than in the previous situations.This alteration of
the baseline causes economic inefficiency to the SO. A mechanism design
should be designed in order to manage properly the signal $r$ to
face this problem when the DR program is based on baseline method.

Lastly, in fig. \ref{fig:incentive summary} are shown the optimal
consumption at the time $t-1$ and $t$, the net consumption ($q_{t}+q_{t-1}$)
and the expected value of consumer profit versus the incentive payment
$p_{2}$. The optimal decision at the setting time is to increase
the consumption as the incentive is raising, namely, the user alters
the baseline in order to improve his profit. Note that whether $p_{2}>0.26\$/kWh$
the energy expenditure is saturated to $q_{max}=20kWh$. In addition,
the rational choice at the period $t$ is to diminish the energy consumption
to receive the benefits of participating in the PTR program. For $p_{2}>0.4\$/kWh$,
the consumed energy goes to zero. Besides, whether $p_{2}\in\left[0,0.26\right)$ $\$/kWh$,
the net consumption is less or equal to $16kWh$, that is, the user
shifts his energy consumption. On the other hand, for $p_{2}>0.26\$/kWh$,
the user spends more energy that he needs, taking into account all
periods. Finally, the expected value of consumer profit is an increasing
function, thus, the incentive payment improve the consumer benefits.
However, the PTR mechanism is favorable for the SO as long as $p_{2}<p$
because the consumer is shifting his energy consumption. In other
situations, the incentive goes against with the objectives of a DR program.

\subsection{Uncertainty variation}

\begin{figure}[H]
\begin{minipage}[c][1\totalheight][t]{0.45\textwidth}%
\begin{center}
\textbf{\Huge{}\includegraphics[scale=0.4]{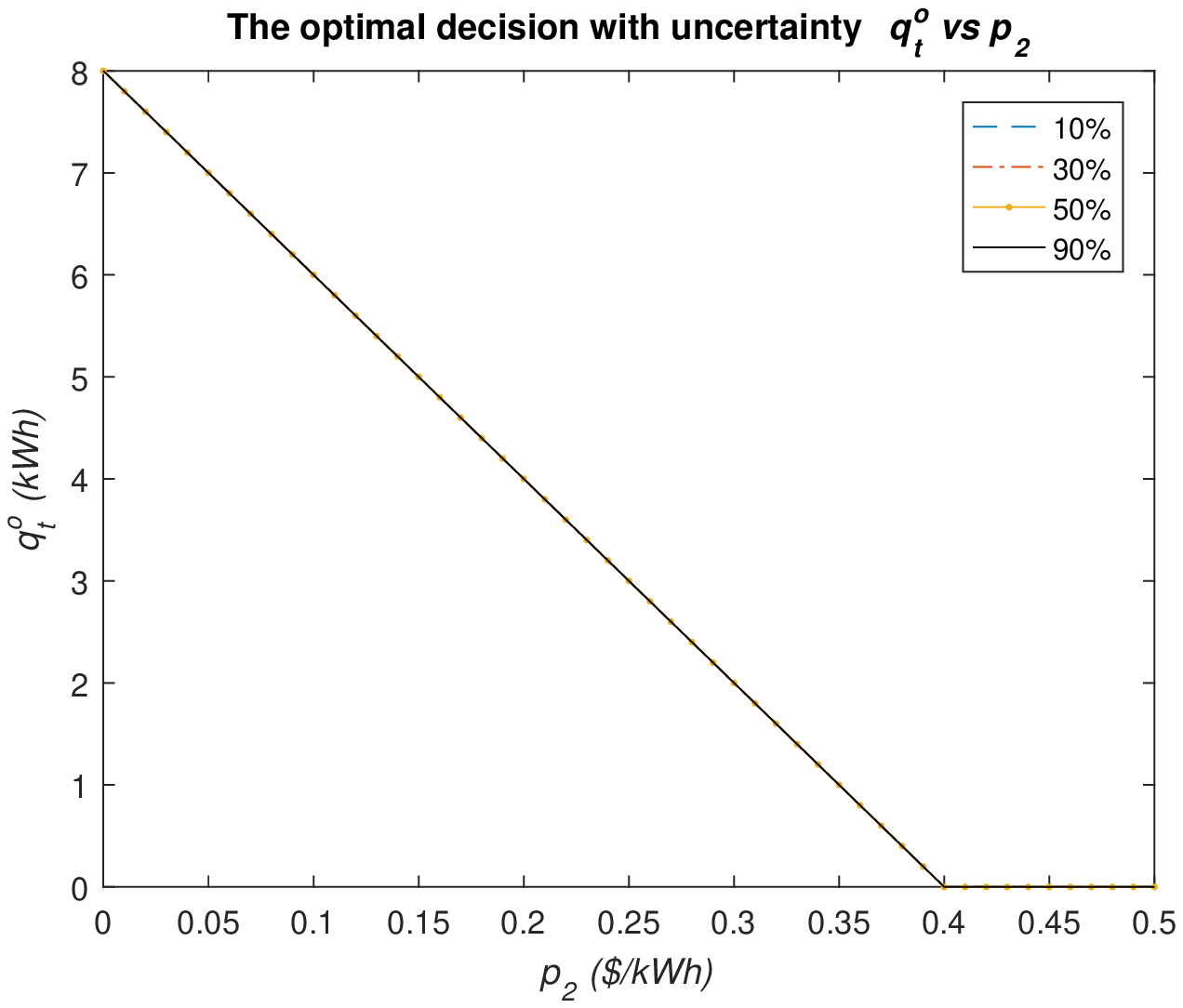}}
\par\end{center}{\Huge \par}

\caption{\label{fig:qt}Optimal decision for $q_{t}$ with uncertainty of 10\%,
30\%, 50\% and 90\% according to the incentive.}
\end{minipage}\hfill{}%
\begin{minipage}[c][1\totalheight][t]{0.45\textwidth}%
\begin{center}
\includegraphics[scale=0.4]{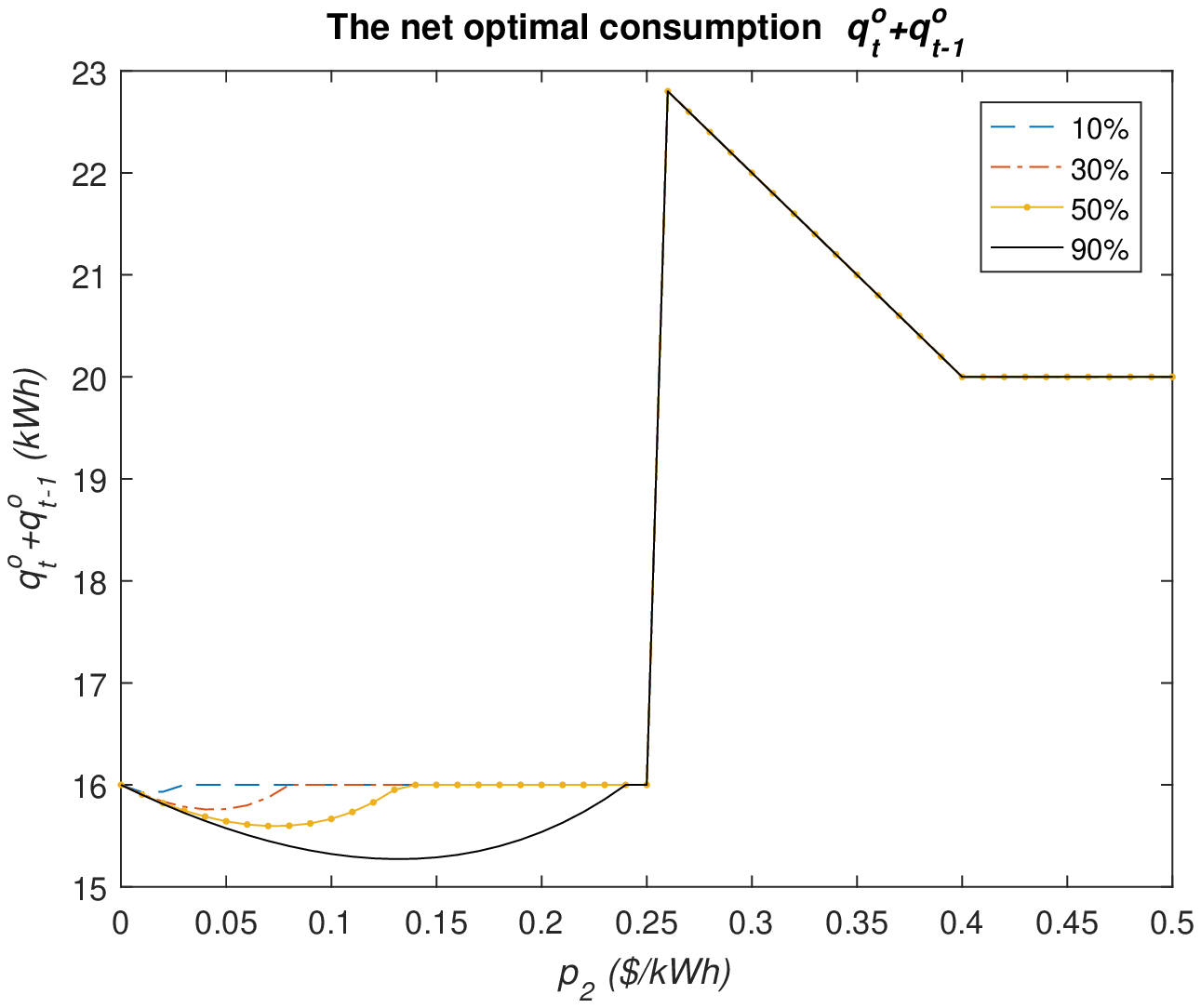}
\par\end{center}

\caption{\label{fig:Net}Net consumption with uncertainty of 10\%, 30\%, 50\%
and 90\% according to the incentive.}
\end{minipage}

\begin{minipage}[c][1\totalheight][t]{0.45\textwidth}%
\begin{center}
\includegraphics[scale=0.4]{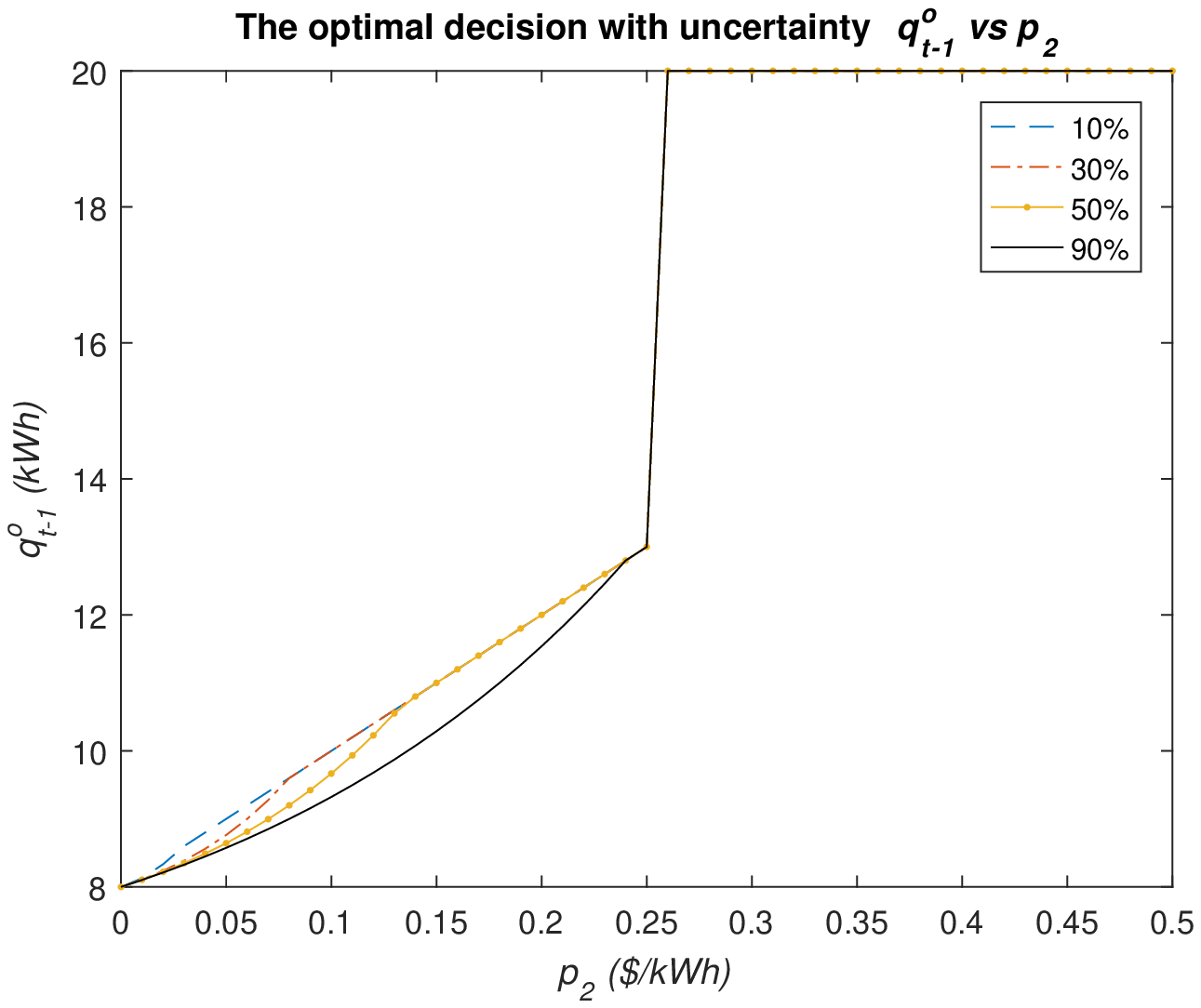}
\par\end{center}

\caption{\label{fig:q_t_1}Optimal decision for $q_{t-1}$ with uncertainty
of 10\%, 30\%, 50\% and 90\% according to the incentive.}
\end{minipage}\hfill{}%
\begin{minipage}[c][1\totalheight][t]{0.45\textwidth}%
\begin{center}
\includegraphics[scale=0.4]{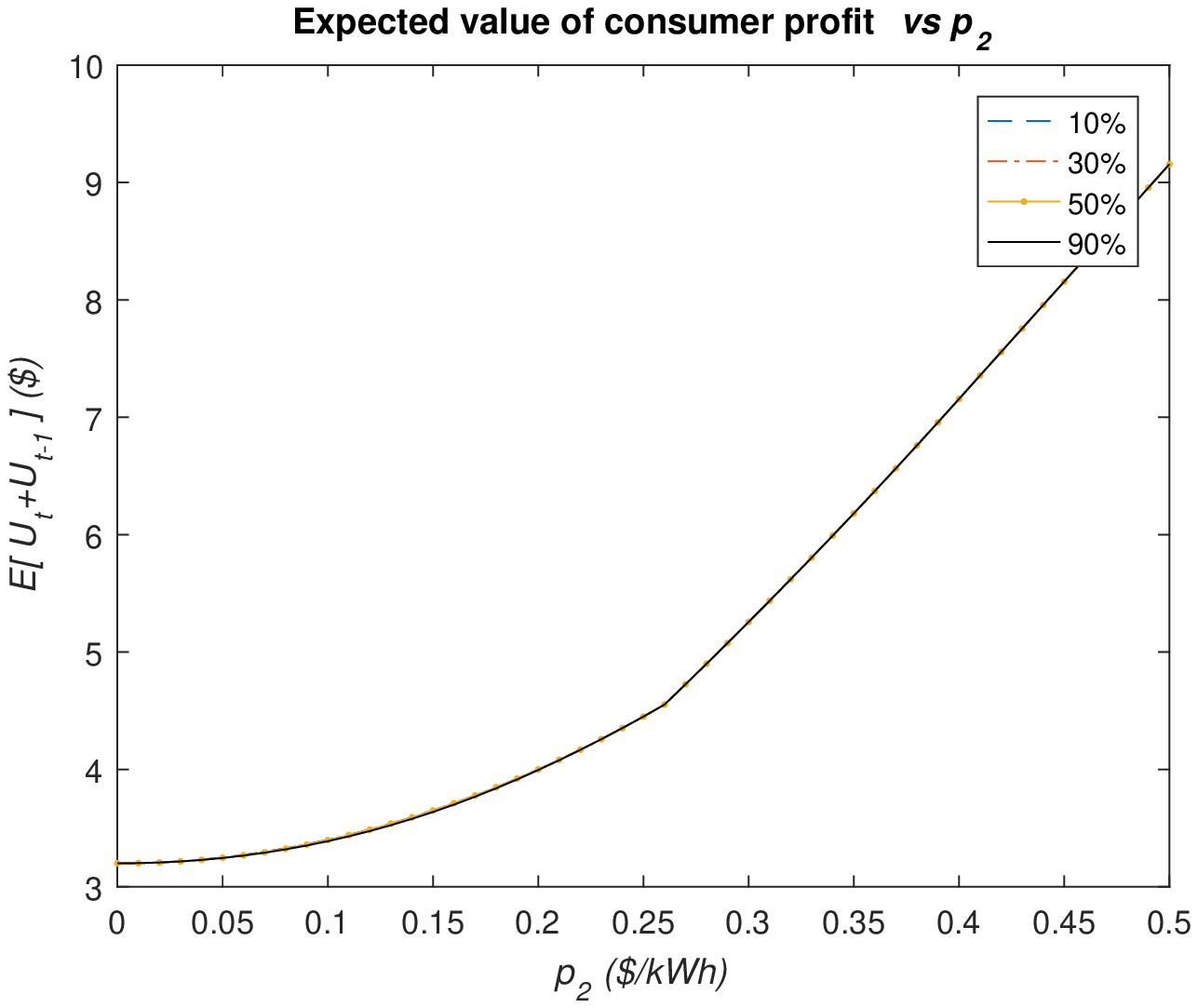}
\par\end{center}

\caption{\label{fig:Expected}Expected value for $U_{t-1}+U_{t}$ with uncertainty
of 10\%, 30\%, 50\% and 90\% according to the incentive.}
\end{minipage}
\end{figure}

In this part, user behaviors for different uncertainty levels are
analyzed. The realization of uniform random variable $\theta_{t}$
is settled with four different supports in order to assess the uncertainty.
These supports are proposed as percentages of the deterministic baseline
$\overline{q}$. For this survey are considered the following percentages:
10\%, 30\%, 50\% and 90\%. In fig \ref{fig:qt} is compared the optimal
decision $q_{t}$ for all the stated uncertainties. Note that the
optimal choice at time $t$ does not depend on the uncertainty level
owing to in this period. Moreover, in fig \ref{fig:q_t_1} is shown
the rational choices at the period $t-1$. For $p_{2}\in\left[0,0.26\right)$,
the user with high uncertainty (e.g. with 50\%) should spend less
energy than a predictable consumer (e.g. with 10\%) since his consumption
is unknown then he reduces his consumption for facing this variation
and pursuing the benefits of the PTR program.Whether the incentive
is greater than the retail price, hence, all decisions are saturated.
Furthermore, A similar behavior is found whether the net consumption
is analyzed (see fig. \ref{fig:Net}). It is vital to restate that
the consumption variation is perceived when the incentive is lower
than the retail price. Lastly, In fig. \ref{fig:Expected} is presented
the expected value of consumer profit. The expected profits are the
same for all percentages because each situation has the same preferences.
In brief, the uncertainty affects low payments of incentive, therefore,
the user does not have certainty related to his consumption pattern
under this conditions, then, his best strategy is to be cautious and
spend less energy than a predictable consumer. 

\begin{figure}[H]
	\begin{centering}
		\includegraphics[scale=0.7]{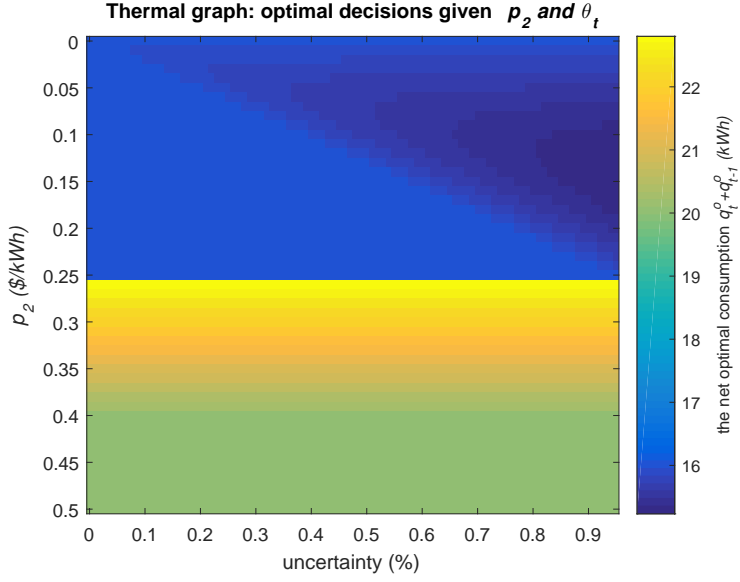}
		\par\end{centering}
	
	\caption{\label{fig:thermal}Thermal graph of optimal decisions given the incentive and uncertainty variation.}
\end{figure}

Finally, in fig \ref{fig:thermal} is shown a thermal graph of the net optimal consumption according to the incentive price $p_2$ and uncertainty variation $\theta_t$ as a plot summary. An important threshold is when the incentive is equal to the retail price, i.e., $p_2= 0.26\$/kWh$ . Even more, the maximum  consumption is detected  when $p_2$ is just slightly higher than $p_2$, rising around $22$  $kWh$ represented by a yellow color. In this situation, the optimal consumption does not change with the uncertainty level. In addition, the rational consumption decreases for incentives between 0.26$\$/kWh$ to 0.4$\$/kWh$. For higher incentives, the net consumption remains constant in $20 kWh$. On the other hand, when the incentive is lower than the retail price, the optimal consumption depends on uncertainty variation. If a user is not sure of his demand then the optimal choice is to consume less energy than a predictable consumer. In particular, this non-linear pattern is depicted by variations in blue tones of fig \ref{fig:thermal}. Furthermore, the maximum energy consumption is $16$ $kWh$ for $p_2$ lower than $p$, therefore, a rational consumer shifts or reduces his load requirement under this incentive conditions.     

\section{Conclusions }

In this paper was analyzed the rational behavior of a consumer that participates in a PTR program within an electricity market. The problem was addressed using a stochastic programming algorithm. A closed-form solution was found for a two-periods framework. The previous consumption was taken as the baseline and it was assumed that the user is always called to participate in the PTR program. The formulation allowed linking the consumer decisions among different consumption periods. Furthermore, uncertainty in load requirements was considered and coupled through conditional expectation. 

It was found that a rational user changes his consumption pattern in order to alter the baseline construction and increase his well-being. Whether the incentive is lower than the regular energy price, the user's best strategy is to shift the energy consumption from the DR event to the baseline settling period. Otherwise, whether the incentive is greater than the retail price then the consumer maximizes his profits consuming as much energy as possible during the baseline setting period, harming the system reliability. In addition, the effect of uncertainty in the consumer energy requirement was analyzed. It was found that the best decision for a consumer with high uncertainty is to spend less energy than a predictable user. 

PTR programs aim to induce users to reduce their energy consumption during a peak event. However, the analysis of the proposed model showed that in most cases, users shift or increase their demand in order to maximize their profits. Only those consumers with high levels of uncertainty reduce their consumption when the incentive is lower than the retail price.  Therefore, it was found that a PTR program is not suitable if the SO is seeking a net reduction of energy consumption on the demand side.

For future works, a mechanism design for the demand side would be a significant improvement for  DR programs based on baseline methods. This mechanism should include a participation condition or what should be the minimum incentive in order to motivate energy reduction or shifting, this property is known as individual rationality constraint. Furthermore, incentive-compatible, budget balance and efficiency should be evaluated for this kind of incentive-based demand response programs. Moreover,  the key solution is associated with controlling properly the user participation. Therefore, it could be interesting to discuss what kind of technology is required to follow the consumer behavior
under this program in order to ensure system efficiency.

\section*{Acknowledgements}
J. Vuelvas received a doctoral scholarship from COLCIENCIAS (Call 647-2014). This work has been partially supported by COLCIENCIAS (Grant 1203-669-4538, Acceso Universal a la Electricidad).

\appendix

\section{Proof of the Theorem \ref{thm:The-optimal-consumption qt}}
\begin{proof}
The optimization problem is analyzed by intervals according to the
established setting. Then the global maximum is found.

\emph{Strategy A1:} $r=1$ (Called), $\boldsymbol{q_{t}^{o}}\geq q_{t-1}$
(Non-participant) and $0\leq\boldsymbol{q_{t}^{o}}\leq q_{t}^{*}+\frac{p}{\gamma}$
($G$ non-saturated)

\[
\left[\boldsymbol{q_{t}^{o}}\right]=\mathrm{argmax}_{q_{t}\in\left[0,q_{t}^{*}+\frac{p}{\gamma}\right]}\;-\frac{\gamma}{2}\left(q_{t}-q_{t}^{*}\right)^{2}+p\left(q_{t}-q_{t}^{*}\right)+k-pq_{t}
\]

The first-order optimality condition yields to 
\begin{equation}
\boldsymbol{q_{t}^{0}}=q_{t}^{*}=\overline{q}+\theta_{t}\label{eq:ut_r=00003D00003D00003D1_noreduce}
\end{equation}

\emph{Strategy A2:} $r=1$ (Called), $\boldsymbol{q_{t}^{o}}\geq q_{t-1}$(Non-participant)
and $\boldsymbol{q_{t}^{o}}>q_{t}^{*}+\frac{p}{\gamma}$ ($G$ saturated).

\[
\left[\boldsymbol{q_{t}^{o}}\right]=\mathrm{argmax}_{q_{t}\in\left[q_{t}^{*}+\frac{p}{\gamma},q_{max}\right]}\;-\frac{p^{2}}{2\gamma}+\frac{p^{2}}{\gamma}+k-pq_{t}
\]

This function is unbounded below. The corner solution is

\begin{equation}
\boldsymbol{q_{t}^{o}}=q_{t}^{*}+\frac{p}{\gamma}\label{eq:caseA2}
\end{equation}

Comparing the optimal solutions (\ref{eq:ut_r=00003D00003D00003D1_noreduce})
and (\ref{eq:caseA2}), the optimal strategy is (\ref{eq:ut_r=00003D00003D00003D1_noreduce})
when the user is called but does not participate in DR.

\emph{Strategy B1:} $r=1$ (Called), $\boldsymbol{q_{t}^{o}}<q_{t-1}$
(Participant), and $0\leq\boldsymbol{q_{t}^{o}}\leq q_{t}^{*}+\frac{p}{\gamma}$
($G$ non-saturated).

\[
\left[\boldsymbol{q_{t}^{o}}\right]=\mathrm{argmax}_{q_{t}\in\left[0,q_{t}^{*}+\frac{p}{\gamma}\right]}\;-\frac{\gamma}{2}\left(q_{t}-q_{t}^{*}\right)^{2}+p\left(q_{t}-q_{t}^{*}\right)+k-pq_{t}+p_{2}\left(q_{t-1}-q_{t}\right)
\]

The first-order optimality condition yields to 
\begin{equation}
\boldsymbol{q_{t}^{0}}=q_{t}^{*}-\frac{p_{2}}{\gamma}=\overline{q}+\theta_{t}-\frac{p_{2}}{\gamma}\label{eq:ut_r=00003D00003D00003D1sireduce}
\end{equation}

\emph{Strategy B2:} $r=1$ (Called), $\boldsymbol{q_{t}^{o}}<q_{t-1}$
(Participant) and $\boldsymbol{q_{t}^{o}}>q_{t}^{*}+\frac{p}{\gamma}$
(Saturated).

\[
\left[\boldsymbol{q_{t}^{o}}\right]=\mathrm{argmax}_{q_{t}\in\left[q_{t}^{*}+\frac{p}{\gamma},q_{max}\right]}\;-\frac{p^{2}}{2\gamma}+\frac{p^{2}}{\gamma}+k-pq_{t}+p_{2}\left(q_{t-1}-q_{t}\right)
\]

This function is unbounded below. The corner solution is

\begin{equation}
\boldsymbol{q_{t}^{o}}=q_{t}^{*}+\frac{p}{\gamma}\label{eq:caseB2}
\end{equation}

Comparing the optimal solutions (\ref{eq:ut_r=00003D00003D00003D1sireduce})
and (\ref{eq:caseB2}), the optimal strategy is (\ref{eq:ut_r=00003D00003D00003D1sireduce})
when the user is called and participates in DR.

\emph{Strategy C:} Importantly, the incentive $p_{2}$ can be so high
to drive (\ref{eq:ut_r=00003D00003D00003D1sireduce}) negative values.
As there is no sense in a negative consumption, the problem is limited
to $\left[0,q_{max}\right]$, then

\begin{equation}
\boldsymbol{q_{t}^{o}}=0\;\;\text{if}\;\;\theta_{t}\leq\frac{p_{2}}{\gamma}-\overline{q}\label{eq:0}
\end{equation}

\emph{Strategy D1:} $r=0$ (Non-called) and $0\leq\boldsymbol{q_{t}^{o}}\leq q_{t}^{*}+\frac{p}{\gamma}$
(Non-saturated)

\[
\left[\boldsymbol{q_{t}^{o}}\right]=\mathrm{argmax}_{q_{t}\in\left[0,q_{t}^{*}+\frac{p}{\gamma}\right]}\;-\frac{\gamma}{2}\left(q_{t}-q_{t}^{*}\right)^{2}+p\left(q_{t}-q_{t}^{*}\right)+k-pq_{t}
\]

The first-order optimality condition yields to

\begin{equation}
\boldsymbol{q_{t}^{0}}=q_{t}^{*}=\overline{q}+\theta_{t}\label{eq:ut_r=00003D00003D00003D0}
\end{equation}

\emph{Strategy D2:} $r=0$ (Non-called) and $\boldsymbol{q_{t}^{o}}>q_{t}^{*}+\frac{p}{\gamma}$
(saturated)

\[
\left[\boldsymbol{q_{t}^{o}}\right]=\mathrm{argmax}_{q_{t}\in\left[q_{t}^{*}+\frac{p}{\gamma},q_{max}\right]}\;-\frac{p^{2}}{2\gamma}+\frac{p^{2}}{\gamma}+k-pq_{t}
\]

This function is unbounded below. The corner solution is

\begin{equation}
\boldsymbol{q_{t}^{o}}=q_{t}^{*}+\frac{p}{\gamma}\label{eq:solerrada}
\end{equation}

Comparing (\ref{eq:ut_r=00003D00003D00003D0}) and (\ref{eq:solerrada}),
the optimal strategy when the user is not called is (\ref{eq:ut_r=00003D00003D00003D0}).

Note that, when called ($r=1$), the user decides to participate (\emph{Strategy
B}) when $\boldsymbol{q_{t}^{o}}<q_{t-1}$, i.e., $\theta_{t}<q_{t-1}-\overline{q}+\frac{p_{2}}{\gamma}$.
While the user does not participate (\emph{Strategy A}) when $\boldsymbol{q_{t}^{o}}>q_{t-1}$,
i.e., $\theta_{t}>q_{t-1}-\overline{q}$. Then, for any realization
of the additive uncertainty $\theta_{t}$ within the interval $q_{t-1}-\overline{q}<\theta_{t}<q_{t-1}-\overline{q}+\frac{p_{2}}{\gamma}$,
there are two local maxima.

In order to find the global solution, the payoff in strategies A and B
are compared. The critical value of $\theta_{t}$ that provides the
same payoff in both strategies is:{\footnotesize{} }
\begin{equation}
U\left(\overline{q}+\theta_{t}-\frac{p_{2}}{\gamma},\theta_{t},q_{t-1}\right)=U\left(\overline{q}+\theta_{t},\theta_{t},q_{t-1}\right)\label{eq:critico}
\end{equation}

Solving for $\theta_{t}$,

\begin{equation}
\theta_{t}=q_{t-1}-\overline{q}+\frac{p_{2}}{2\gamma}\label{eq:condicion}
\end{equation}
Eq. (\ref{eq:condicion}) gives the limit of the uncertain load when
the user commutes from strategy A to strategy B.

Organizing by intervals the results (\ref{eq:ut_r=00003D00003D00003D0}),
(\ref{eq:ut_r=00003D00003D00003D1_noreduce}), (\ref{eq:ut_r=00003D00003D00003D1sireduce})
and (\ref{eq:0}), the solution is given by theorem \ref{thm:The-optimal-consumption qt}. 
\end{proof}

\section{Proof of the Theorem \ref{th case 1}}
\begin{proof}
Let $r=1$, i.e., the user is always called to participate in the
PTR program. Under the assumption (see the corollary \ref{cor:The-expected-value}
or the fig. \ref{fig:Flowchart}) that $\underline{\theta}>\frac{p_{2}}{\gamma}-\overline{q}$,
namely, strategy $C$ does not exist in the density probability function
$f_{\theta}\left(\theta_{t}\right)$ (See fig. \ref{fig:Probability-density-function}
with zero probability for strategy $C$). Also, it is assumed that
$\overline{\theta}=-\underline{\theta}$ and $\underline{\theta}>\frac{p_{2}}{\gamma}-\overline{q}$
and the parameters $\overline{q}$ and $p_{2}$ are positives.

Subsequently, the net payoff function for all periods is given by
figure \ref{fig:low uncertainty}. Notice that the intervarls are
given when the conditions $q_{t-1}-\overline{q}+\frac{p_{2}}{2\gamma}$
is equal to $\underline{\theta}$ and $\overline{\theta}$. Besides,
the saturation part according the utility function (equation (\ref{eq:g}))
is assumed between $\overline{q}+\overline{\theta}-\frac{p_{2}}{2\gamma}\leq\overline{q}+\frac{p}{\gamma}\leq q_{max}$.
This assumption about the saturation part is motivated due to $\overline{\theta}$
is relatively small when the user has not too much uncertainty. Thus,
an optimization problem is formulated by intervals according to strategies
that are feasible. Then, the maximum global is found comparing all
the local maxima. 

\begin{figure}[H]
\begin{centering}
\includegraphics{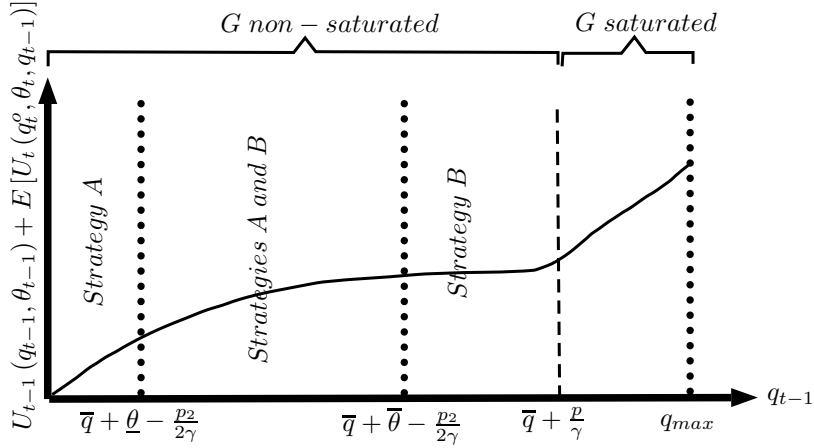}
\par\end{centering}

\caption{\label{fig:low uncertainty}Net payoff function when strategy $C$
does not exist.}
\end{figure}

The first local maximum is found when strategy A is feasible. 

\[
\begin{array}{c}
\left[\boldsymbol{q_{t-1}^{o}}\right]=\mathrm{arg}\mathrm{max}_{q_{t-1}}\;G\left(q_{t-1}-\theta_{t-1}\right)-pq_{t-1}+E_{A}\\
\mathrm{s.t.}\quad0\leq q_{t-1}\leq\overline{q}+\underline{\theta}-\frac{p_{2}}{2\gamma}
\end{array}
\]

The Karush Kuhn Tucker conditions for the above formulations are:

\[
\begin{array}{c}
\frac{\partial}{\partial q_{t-1}}\left(G\left(q_{t-1}-\theta_{t-1}\right)-pq_{t-1}+E_{A}\right)+\mu_{1}-\mu_{2}=0\\
0\leq q_{t-1}\bot\mu_{1}\geq0\\
q_{t-1}\leq\overline{q}+\underline{\theta}-\frac{p_{2}}{2\gamma}\perp\mu_{2}\geq0
\end{array}
\]

Being the $E\left[\theta_{t-1}\right]=0$, it is found that: 

{\scriptsize{}
\[
E\left[\boldsymbol{q_{t-1}^{o}(\theta_{t-1})}\mid\underline{\theta}>\frac{p_{2}}{\gamma}-\overline{q}\,and\,q_{t-1}\in\left[0,\overline{q}+\underline{\theta}-\frac{p_{2}}{2\gamma}\right]\right]=\left\{ \begin{array}{cc}
0 & -\overline{q}>0\\
\overline{q} & \overline{q}\geq0\:and\,p_{2}<2\gamma\underline{\theta}\\
\overline{q}+\underline{\theta}-\frac{p_{2}}{2\gamma} & p_{2}\geq2\gamma\underline{\theta}
\end{array}\right.
\]
}{\scriptsize \par}

Therefore the unique feseable solution for this situation is:

\begin{equation}
E\left[\boldsymbol{q_{t-1}^{o}(\theta_{t-1})}\mid\underline{\theta}>\frac{p_{2}}{\gamma}-\overline{q}\,and\,q_{t-1}\in\left[0,\overline{q}+\underline{\theta}-\frac{p_{2}}{2\gamma}\right]\right]=\overline{q}+\underline{\theta}-\frac{p_{2}}{2\gamma}\qquad p_{2}\geq0\label{eq:th2_1}
\end{equation}

In addition, The sufficient condition is guaranteed, i.e., $-\gamma<0$

Next, local maxima when strategies $A$ and $B$ are feasible is found
solving the following optimization problem:

\[
\begin{array}{c}
\left[\boldsymbol{q_{t-1}^{o}}\right]=\mathrm{argmax}_{q_{t-1}}\;G\left(q_{t-1}-\theta_{t-1}\right)-pq_{t-1}+E_{AB}\\
\mathrm{s.t.}\quad\overline{q}+\underline{\theta}-\frac{p_{2}}{2\gamma}\leq q_{t-1}\leq\overline{q}+\overline{\theta}-\frac{p_{2}}{2\gamma}
\end{array}
\]

A similar analysis using KKT conditions yields the following result:

{\scriptsize{}
\[
E\left[\boldsymbol{q_{t-1}^{o}(\theta_{t-1})}\mid\underline{\theta}>\frac{p_{2}}{\gamma}-\overline{q}\,and\,q_{t-1}\in\left[\overline{q}+\underline{\theta}-\frac{p_{2}}{2\gamma},\overline{q}+\overline{\theta}-\frac{p_{2}}{2\gamma}\right]\right]=
\]
}{\scriptsize \par}

{\scriptsize{}
\[
\qquad\left\{ \begin{array}{cc}
\overline{q}+\underline{\theta}-\frac{p_{2}}{2\gamma} & p_{2}<2\underline{\theta}\gamma\\
\overline{q}-\frac{p_{2}}{2\gamma}+\frac{p_{2}\left(\overline{\theta}-3\underline{\theta}\right)}{2\left(\left(\overline{\theta}-\underline{\theta}\right)\gamma-p_{2}\right)} & \:2\underline{\theta}\gamma\leq p_{2}<\frac{2}{3}\overline{\theta}\gamma\\
\overline{q}+\overline{\theta}-\frac{p_{2}}{2\gamma} & p_{2}\geq\frac{2}{3}\overline{\theta}\gamma
\end{array}\right.
\]
}{\scriptsize \par}

Then as well, $p_{2}$ is positive, resulting

{\scriptsize{}
\begin{equation}
\begin{array}{c}
E\left[\boldsymbol{q_{t-1}^{o}(\theta_{t-1})}\mid\underline{\theta}>\frac{p_{2}}{\gamma}-\overline{q}\,and\,q_{t-1}\in\left[\overline{q}+\underline{\theta}-\frac{p_{2}}{2\gamma},\overline{q}+\overline{\theta}-\frac{p_{2}}{2\gamma}\right]\right]=\\
\qquad\left\{ \begin{array}{cc}
\overline{q}-\frac{p_{2}}{2\gamma}+\frac{p_{2}\left(\overline{\theta}-3\underline{\theta}\right)}{2\left(\left(\overline{\theta}-\underline{\theta}\right)\gamma-p_{2}\right)} & \:0\leq p_{2}<\frac{2}{3}\overline{\theta}\gamma\\
\overline{q}+\overline{\theta}-\frac{p_{2}}{2\gamma} & p_{2}\geq\frac{2}{3}\overline{\theta}\gamma
\end{array}\right.
\end{array}\label{eq:th2_2}
\end{equation}
}{\scriptsize \par}

However, the sufficient condition is met when $p_{2}<\gamma\left(\overline{\theta}-\underline{\theta}\right)$.
In other circumstances, the solution will be a corner. Given that $\gamma\left(\overline{\theta}-\underline{\theta}\right)>\frac{2}{3}\overline{\theta}\gamma$
then the solution is the same.

Finally, the local maximum when strategy $B$ is feasible. 

\[
\begin{array}{c}
\left[\boldsymbol{q_{t-1}^{o}}\right]=\mathrm{argmax}_{q_{t-1}}\;G\left(q_{t-1}-\theta_{t-1}\right)-pq_{t-1}+E_{B}\\
\mathrm{s.t.}\quad\overline{q}+\overline{\theta}-\frac{p_{2}}{2\gamma}\leq q_{t-1}\leq q_{max}
\end{array}
\]

which has the following solution,

{\scriptsize{}
\begin{equation}
E\left[\boldsymbol{q_{t-1}^{o}(\theta_{t-1})}\mid\underline{\theta}>\frac{p_{2}}{\gamma}-\overline{q}\,and\,q_{t-1}\in\left[\overline{q}+\overline{\theta}-\frac{p_{2}}{2\gamma},q_{max}\right]\right]=\left\{ \begin{array}{cc}
\overline{q}+\overline{\theta}-\frac{p_{2}}{2\gamma} & 0\leq p_{2}<\frac{2}{3}\overline{\theta}\gamma\\
\overline{q}+\frac{p_{2}}{\gamma} & \:\frac{2}{3}\overline{\theta}\gamma\leq p_{2}<p\\
q_{max} & p_{2}\geq p
\end{array}\right.\label{eq:th2_3}
\end{equation}
}{\scriptsize \par}

Furthermore, The minimum condition is guaranteed, i.e., $-\gamma<0$.

Lastly, comparing the net payoff at the local maxima given by (\ref{eq:th2_1}),
(\ref{eq:th2_2}) and (\ref{eq:th2_3}). The global solution is:

\[
E\left[\boldsymbol{q_{t-1}^{o}(\theta_{t-1})}\mid\underline{\theta}>\frac{p_{2}}{\gamma}-\overline{q}\right]=\left\{ \begin{array}{cc}
\overline{q}-\frac{p_{2}}{2\gamma}+\frac{2p_{2}\overline{\theta}}{2\overline{\theta}\gamma-p_{2}} & 0\leq p_{2}<\frac{2}{3}\overline{\theta}\gamma\\
\overline{q}+\frac{p_{2}}{\gamma} & \frac{2}{3}\overline{\theta}\gamma\leq p_{2}<p\\
q_{max} & p\leq p_{2}<\gamma\left(\underline{\theta}+\overline{q}\right)
\end{array}\right.
\]

\end{proof}

\section{Proof of the theorem \ref{th case 2}}
\begin{proof}
The mathematical expression $\frac{p_{2}}{\gamma}-\overline{q}$ is
located within the limits of the probability density function (see
fig. \ref{fig:Probability-density-function}). Furthermore, $\overline{q}+\frac{p}{\gamma}>\overline{q}+\frac{p}{\gamma}$,
i.e., the saturated point is when strategies $B$ and $C$ are feasible
as it is shown in fig. \ref{fig:medium uncertainty}. Also, let $\underline{\theta}\leq\frac{p_{2}}{\gamma}-\overline{q}<\overline{\theta}$. 

\begin{figure}[H]
\begin{centering}
\includegraphics{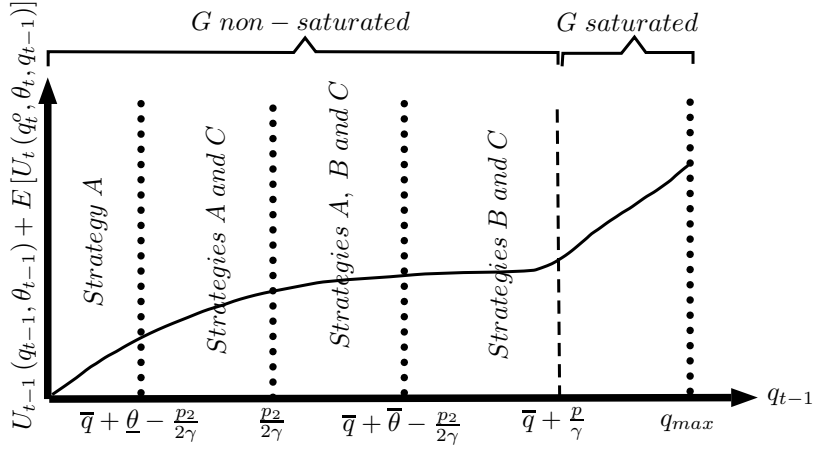}
\par\end{centering}

\caption{\label{fig:medium uncertainty}Net payoff function when strategy $C$
is inside of probability density function.}
\end{figure}

It is uncomplicated to show the following statements

\[
\max_{q_{t-1}}\;G\left(q_{t-1}-\theta_{t-1}\right)-pq_{t-1}+E_{AB}=\max_{q_{t-1}}\;G\left(q_{t-1}-\theta_{t-1}\right)-pq_{t-1}+E_{ABC}
\]

\[
\max_{q_{t-1}}\;G\left(q_{t-1}-\theta_{t-1}\right)-pq_{t-1}+E_{B}=\max_{q_{t-1}}\;G\left(q_{t-1}-\theta_{t-1}\right)-pq_{t-1}+E_{BC}
\]

Therefore, the last two zones ($\left[q_{max},\overline{q}+\overline{\theta}-\frac{p_{2}}{2\gamma}\right]$
and $\left[\overline{q}+\overline{\theta}-\frac{p_{2}}{2\gamma},\frac{p_{2}}{2\gamma}\right]$)
from fig.\ref{fig:low uncertainty} and fig. \ref{fig:medium uncertainty}
have some similarities. Whether the reader follows the same steps
of the proof of the theorem \ref{th case 1} then the solution for
this theorem is: 

\[
E\left[\boldsymbol{q_{t-1}^{o}(\theta_{t-1})}\mid\underline{\theta}<\frac{p_{2}}{\gamma}-\overline{q}<\overline{\theta}\right]=\left\{ \begin{array}{cc}
\overline{q}-\frac{p_{2}}{2\gamma}+\frac{2p_{2}\overline{\theta}}{2\overline{\theta}\gamma-p_{2}} & \gamma\left(\underline{\theta}+\overline{q}\right)\leq p_{2}<\frac{2}{3}\overline{\theta}\gamma\\
\overline{q}+\frac{p_{2}}{\gamma} & \frac{2}{3}\overline{\theta}\gamma\leq p_{2}<p\\
q_{max} & p<p_{2}\leq\gamma\left(\overline{\theta}+\overline{q}\right)
\end{array}\right.
\]

\end{proof}

\section{Proof of the theorem \ref{th case 3}}

\begin{figure}[H]
\begin{centering}
\includegraphics{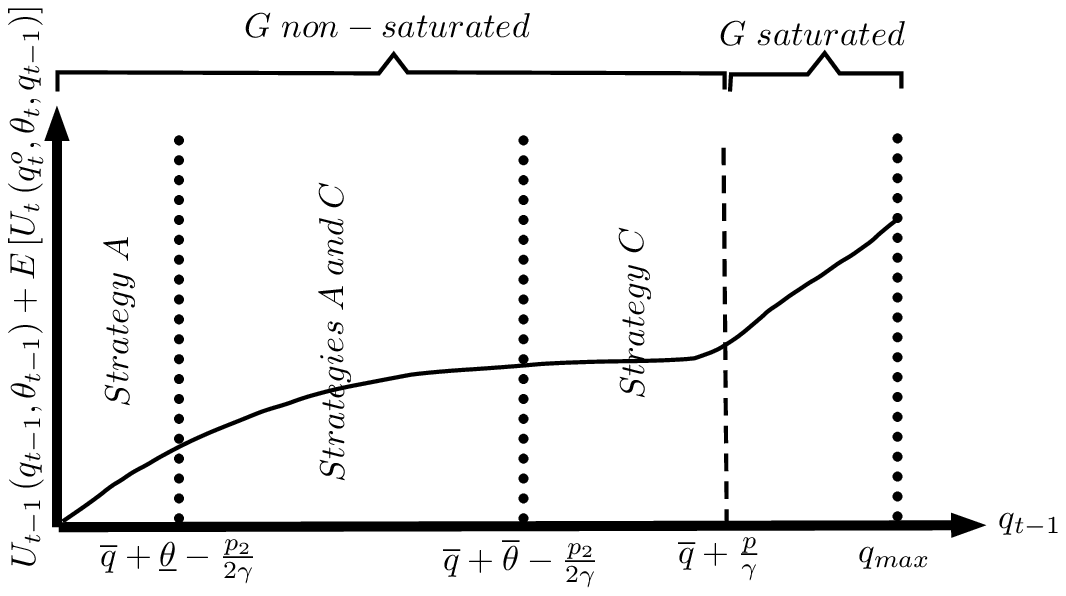}
\par\end{centering}

\caption{\label{fig:High_uncertainty}Net payoff function when strategy $C$
is greater than $\overline{\theta}$.}
\end{figure}

This theorem is proved using the same procedure than theorem (\ref{th case 1})
and (\ref{th case 2}). 

Fig. \ref{fig:High_uncertainty} depicts all the zones feasible for
this case.

\section*{-----------------}

\bibliographystyle{elsarticle-harv}
\addcontentsline{toc}{section}{\refname}\bibliography{PTR_revista}

\end{document}